\documentclass{article}
\usepackage[utf8]{inputenc}
\usepackage[margin=1in]{geometry}
\usepackage{setspace}
\usepackage{amssymb}
\usepackage{amsmath}
\usepackage{amsthm}
\usepackage{bbm}
\usepackage{graphicx}
\usepackage{mwe}
\usepackage{color,soul}
\usepackage{multirow}
\newtheorem{theorem}{Theorem}
\newtheorem{lemma}{Lemma}
\newtheorem{corollary}{Corollary}

\usepackage[algoruled,boxed,lined]{algorithm2e}
\usepackage{natbib}
\usepackage{bm}

\usepackage{hyperref}
\hypersetup{colorlinks,allcolors=black}
\DeclareMathOperator\supp{supp}

\title{EnsembleIV: Creating Instrumental Variables from Ensemble Learners for Robust Statistical Inference}
\author{Gordon Burtch$^1$, Edward McFowland III$^2$, Mochen Yang$^3$, Gediminas Adomavicius$^3$ \\
$^1$ Questrom School of Business, Boston University \\
$^2$ Harvard Business School, Harvard University \\
$^3$ Carlson School of Management, University of Minnesota}
\date{Current Draft: 3/5/2023}

\begin{document}
\doublespacing
\maketitle

\section{Introduction} \label{sec:intro}
Empirical researchers in the social sciences are increasingly leveraging a combination of supervised machine learning and statistical inference in a hybrid, two-phase process. In the first phase, a supervised machine learning model is trained to predict a target outcome based on a set of features. In the second phase, predicted values of the target are used as an independent variable, usually within a regression model, for statistical inference purposes. However, because predictions from the first-phase machine learning model are typically imperfect, prediction errors will manifest as measurement error in the second-phase regression model, leading to estimation biases and threatening the validity of inferences.

This two-phase approach is an example of statistical inference with \textit{generated regressors} \citep{pagan1984econometric,oxley1993econometric}. It has seen growing adoption in a number of social science disciplines, partly due to impressive advances in machine learning techniques that enable the efficient extraction of useful information from large amounts of both structured and unstructured data. To name just a few examples, \cite{cengiz2022seeing} built a boosted tree model to predict which workers were affected by changes to the minimum wage based on workers' demographics and then examined the effect of those changes on workers' labor market outcomes. \cite{moreno2014doing} mined text sentiment from user-generated comments on an online service platform, and subsequently estimated the impact of (predicted) sentiment on buyers' purchasing decisions. \cite{zhang2021makes} used deep learning techniques to measure the quality of property images on Airbnb, and studied how image quality, in turn, affects property demand. \cite{wu2020air} estimated the effect of a geography's (predicted) long-term air pollution exposure on the rate of COVID-19 mortality.

While early work that adopted this two-phase approach has largely ignored the measurement error issue that this approach implies, researchers focused on statistical methods have increasingly attended to it over the last few years \citep[e.g.,][]{yang2018mind,wang2020methods,qiao2021correcting,fong2021machine,yang2022achieving}. Compared to the settings in which measurement error has traditionally been addressed, wherein the errors are unobserved, and estimation biases are therefore hard to address, measurement error deriving from a machine learning model's prediction errors is quite unique, in that additional information is typically available about the error that can be leveraged in bias correction. Because the training and evaluation of a machine learning model typically require a manually collected labeled dataset, where the true values of the target outcome (e.g., text sentiment or image quality) are known, one can directly observe the measurement error in the generated variable on the labeled data and quantify its properties (e.g., distributional characteristics or correlations with other variables of interest). This has enabled the use of some existing error-correction techniques to mitigate estimation biases \citep[e.g., Simulation-Extrapolation (SIMEX)][]{Stefanski1995,Kuchenhoff2006,yang2018mind}, as well as the development of new approaches \citep{qiao2021correcting,fong2021machine,yang2022achieving,wei2022unstructured,allon2023machine}.

In this paper, we develop a novel method that leverages instrumental variables to address the measurement error problem. The method consists of three key ingredients, respectively focusing on the generation, transformation, and selection of instruments. First, we propose to use ensemble learning techniques (e.g., random forest) to build the first-phase machine learning model. Doing so generates a set of individual learners (e.g., individual trees in a random forest) whose predictions can serve as candidate instruments for each other. However, these instruments are ``imperfect" in that they are not guaranteed to satisfy the exclusion condition. Accordingly, second, we propose a technique, based on earlier work by \cite{nevo2012identification}, that transforms candidate instruments to ensure they comply with the exclusion condition. Third, to deal with the potential challenge of weak instruments (i.e., instruments that only barely satisfy the relevance condition), we propose and evaluate several approaches to selecting strong instruments, which are subsequently used in instrumental variable regressions to obtain unbiased estimates. We henceforth refer to this method as \textit{EnsembleIV}.

Under standard and mild regularity conditions, we prove the consistency and asymptotic normality of EnsembleIV estimator, which enable asymptotically valid inference. We also carry out empirical evaluations of EnsembleIV to understand its properties and performance on both synthetic and real-world data. Simulation studies with synthetic data demonstrate EnsembleIV's ability to substantially mitigate estimation biases on machine-learning-generated variables in several common regression models. Further, applying EnsembleIV to a real-world dataset of user-generated content on social media, we illustrate how it can be used together with modern deep-learning techniques.

EnsembleIV represents a novel methodological contribution to the nascent literature on robust statistical inference with machine learning-generated variables. While the measurement error problem and its solutions have been extensively studied in the literature, EnsembleIV stands out as an approach that leverages ensemble machine learning techniques to \textit{create} valid instrumental variables. It also improves upon existing approaches in several important aspects. First, EnsembleIV has a more general theoretical foundation, as the method can accommodate (i) both continuous and binary generated variables and (ii) both independent and correlated measurement errors \citep[also known as classical and non-classical errors in the literature,][]{carroll2006measurement}. In contrast, many existing approaches are more limited, e.g., SIMEX \citep{Stefanski1995,yang2018mind} only addresses classical measurement error and MC-SIMEX \citep{Kuchenhoff2006} works only for binary misclassification. Second, prior approaches that use instrumental variables for bias correction, such as ForestIV \citep{yang2022achieving}, typically rely on the ``diversity" of individual learners within an ensemble \citep[namely, the property that prediction errors from different individual learners should be weakly correlated,][]{Breiman2001} to discover valid instruments. EnsembleIV is much less dependent on this property. Even if the ensemble learning technique does not automatically produce diverse individual learners, and thus naturally valid instruments, EnsembleIV can still \textit{create} valid instruments, leveraging the aforementioned IV transformation technique. This allows EnsembleIV to be applied with common ensemble learning algorithm (e.g., random forest, boosting, stacking), greatly enhancing its practical applicability. Third, as will be shown later, EnsembleIV outperforms benchmarks such as ForestIV and regression calibration in terms of estimation efficiency, producing estimates with smaller standard errors in the same sample.

\section{EnsembleIV: Theory and Algorithm} \label{sec:theory}
In this section, we outline the theoretical components needed to create valid instrumental variables from ensemble learners, based on which we design the EnsembleIV algorithm. We use the notations in Table \ref{table:notations} throughout the paper.

\begin{table}[!tbh]
    \centering 
    \caption{Glossary of Notation} \label{table:notations}
    \begin{tabular}{c|l}
        \hline
        Notation & Definition \\
        \hline
        $Y$ & The dependent variable in second-phase regression. \\
        $X$ & The independent variable in second-phase regression. \\
        $\boldsymbol{W}$ & The exogenous control variables in second-phase regression. \\
        $\varepsilon$ & The idiosyncratic error term in second-phase regression. \\
        $\widehat{X}$ & Predicted values of $X$ (generated by the first-phase machine learning model). \\
        $e$ & Measurement (i.e., prediction) error in $\widehat{X}$. \\
        $Z$ & A candidate instrumental variable. \\
        $D_{\bullet}$ & A particular partition of the data, one of training, testing, labeled, or unlabeled. \\
        $M$ & Total number of individual learners in an ensemble. \\
        $\widehat{X}^{(i)}$ & Predicted values of $X$ produced by individual learner $i \in \{1, \ldots, M\}$. \\
        \hline
    \end{tabular}
\end{table}

\subsection{Instrumental Variable Approach to the Measurement Error Problem} \label{sec:theory_IV}
We first formulate the measurement error problem that arises when a machine-learning-generated variable is incorporated into a regression as an independent covariate, then describe the IV approach to address it. Without loss of generality, consider the following regression equation:\footnote{For generalized linear models (e.g., logistic regressions), the same measurement error problem can be formulated based on the latent variable model \citep{wooldridge2002}, where the latent outcome $Y^*$ is linearly related to independent covariates in the same way as specified in \eqref{eq:regression_true}.} 
\begin{equation}
\label{eq:regression_true}
    Y =  X \beta + \boldsymbol{W \Pi} + \varepsilon
\end{equation}
where the idiosyncratic error term $\varepsilon$ is assumed to satisfy the following two assumptions:
\begin{itemize}
    \item [Assumption I:] $E[\varepsilon| X, \boldsymbol{W}] = 0$;
    \item [Assumption II:] Given $\widehat{X}^{(i)} = X + e^{(i)}$, $E[\varepsilon e^{(i)} | X, \boldsymbol{W}]=0, ~ \forall i \in \{1, \ldots, M\}$.
\end{itemize}

Assumption I ensures that, in the absence of measurement error, we are working with a correctly-specified regression equation free from other sources of endogeneity. This allows us to focus on measurement error as the core identification challenge. Assumption II means that prediction errors associated with each individual learner do not have a direct impact on outcome $Y$ beyond the covariates already included in the regression. This setup follows that employed in previous work that deals with correcting estimation bias from machine-learning-generated covariates \citep[e.g.,][]{yang2018mind,qiao2021correcting,fong2021machine,yang2022achieving,allon2023machine,zhang2023debiasing}.

Under the setting that we consider, $X$ can only be observed in a relatively small set of labeled data ($D_{label}$) but is unobserved in the much larger unlabeled data ($D_{unlabel}$, and $|D_{unlabel}| \gg |D_{label}|$). A typical reason for the size disparity between labeled and unlabeled data is the cost of labeling $X$. As an example, in studies that investigate the relationships between textual sentiment and certain dependent variables of interest \citep[e.g.,][]{tirunillai2012does,goh2013social,moreno2014doing}, researchers often need to manually label the sentiment (e.g., by hiring crowd workers, such as Amazon Mechanical Turkers, who read the text and assign sentiment labels).\footnote{We acknowledge that such manual labeling processes can potentially introduce inter-labeler disagreement or error into the labeled data. This can be especially challenging for subjective or open-ended labeling tasks, where different labelers may have different subjective beliefs about ``ground truth" (e.g., labeling ``beauty" based on the image of a person). There are strategies that researchers can adopt to mitigate this type of error, e.g., adding more labelers for each data instance, refining labeling instructions, measuring inter-rater reliability among labelers, and resolving discrepancies via discussion \citep{hopkins2007extracting}. Here, we focus strictly on the prediction error from machine learning models, as labeling error is fundamentally different -- labeling disagreement tends to reflect uncertainty in the labeling process whereas prediction error represents imperfect machine learning predictions.} Doing so for a large volume of text can be very cost-prohibitive. Of course, one can directly estimate regression equation \eqref{eq:regression_true} using $D_{label}$; however, the limited size of $D_{label}$ may result in imprecise estimates, potentially impairing subsequent statistical inference and the ultimate managerial or policy decision. To make use of the larger $D_{unlabel}$, one increasingly common approach has been to build machine learning models on $D_{label}$ to predict the values of $X$ on $D_{unlabel}$.

However, the predicted values $\widehat{X}$ generally contain some degree of prediction error, defined as $e := \widehat{X} - X$. When $\widehat{X}$ is added into regression equation \eqref{eq:regression_true} as a surrogate for $X$, the prediction errors become measurement errors, and the regression that is actually estimated can be written as:
\begin{equation}
\label{eq:regression_est}
    Y = \widehat{X} \beta + \boldsymbol{W \Pi} + (\varepsilon - e \beta)
\end{equation}

We denote $u := \varepsilon - e\beta$ as the error term in the estimated regression, and $Cov(\widehat{X}, u) = Cov(X + e, \varepsilon - e\beta) = -\beta (Cov(X,e) + Var(e))$. Aside from a very unlikely scenario wherein $Cov(X,e)$ precisely cancels out $Var(e)$, the regression will suffer from endogeneity and produce biased and inconsistent estimates.

It is important to note that we do not make assumptions regarding the distributions of $X$ or $\widehat{X}$ -- they can be continuous or binary -- nor do we restrict the relationship between $e$ and $X$ (i.e., $e$ could be independent of, or correlated with, $X$). In other words, our proposed approach is applicable for both continuous and binary mismeasured covariates, and under both independent and correlated measurement error.

A standard solution to the measurement error problem is the instrumental variable approach \citep{buzas1996instrumental,greene2003econometric,hu2008instrumental}. A valid instrumental variable (IV), $Z$, needs to satisfy two conditions, namely (i) \textit{relevance}, $Cov(Z, \widehat{X}) \neq 0$, i.e., the IV should be correlated with the mismeasured covariate; and (ii) \textit{exclusion}, $Cov(Z, u) = -\beta Cov(Z, e) = 0$, i.e., the IV should be uncorrelated with the measurement error. If such an IV can be identified, one can obtain consistent estimates via IV regression, e.g., two-stage least-squares \citep[2SLS,][]{greene2003econometric,wooldridge2002}. IV regression is known to be consistent in the limit and significantly mitigates estimation bias in (large) finite samples.

Despite the attractive theoretical properties of the IV approach, in practice, it is often challenging to identify instruments that satisfy the exclusion condition because the measurement error (or, more generally, the source of endogeneity) is not directly observable. As such, the exclusion condition is typically argued conceptually \citep{conley2012plausibly}. Again, in our context, this is not the case; measurement error \textit{can} be observed directly in the labeled sample of data that is used to train and test the machine learning model.

\subsection{Ensemble Learners as Instrumental Variables} \label{sec:theory_ensemble}
We now describe the key theoretical components of EnsembleIV. Suppose the first-phase machine learning model is an ensemble model with $M$ individual learners (e.g., a random forest with $M$ trees). Instead of the common practice of using the aggregated predictions, $\widehat{X} = \frac{1}{M} \sum_i \widehat{X}^{(i)}$, in the second-phase regression, we consider using the predictions from one individual learner, $\widehat{X}^{(i)}$, employing those predictions as a mismeasured variable, and relying on the predictions from other individual learners, $\widehat{X}^{(j)} (j \neq i)$, as candidate instruments.\footnote{In numeric prediction tasks, $\widehat{X}^{(i)}$ is comprised of the raw numerical predictions of the $i$-th learner. In binary classification tasks, it is the probability prediction of the $i$-th learner.} 

Because the different individual learners are all predictive of the target ground truth to some extent, their predictions will generally be correlated with one another, i.e., $Cov(\widehat{X}^{(j)}, \widehat{X}^{(i)}) \neq 0$. This satisfies the relevance condition for valid IVs. However, there is no theoretical guarantee that these candidate IVs will satisfy the exclusion condition, which requires that $Cov(\widehat{X}^{(j)}, \widehat{X}^{(i)} - X) = 0$. To address the exclusion violation, we apply an IV transformation technique developed based on prior work by \cite{nevo2012identification}.

Specifically, suppose a candidate IV, $Z$, is ``imperfect" in the sense that it violates the exclusion condition, i.e., $Cov(Z,u) = -\beta Cov(Z,e) \neq 0$. \cite{nevo2012identification}, relaxing the assumption of strict exclusion, allow an instrumental variable to be correlated with the error term, and apply the weaker assumptions that (i) the correlation between the instrumental variable and the error term has the same sign as the correlation between the endogenous regressor and the error term and (ii) that the instrumental variable is less correlated with the error term than is the endogenous regressor. They derive analytic bounds for the parameters and demonstrate the application of their method to produce set estimates.

Leveraging portions of their theoretical observations and the fact that we observe measurement error directly in our setting, we develop a procedure to transform a candidate IV so that it satisfies the exclusion restriction. Denote correlation coefficient $\rho_{Zu} = \frac{Cov(Z,u)}{\sigma_Z \sigma_u}$ and $\rho_{\widehat{X}u} = \frac{Cov(\widehat{X},u)}{\sigma_{\widehat{X}} \sigma_u}$ and consider the following quantity:
\begin{equation}
\label{eq:lambda}
    \lambda = \frac{\rho_{Zu}}{\rho_{\widehat{X}u}} = \frac{Cov(Z,u)}{Cov(\widehat{X},u)} \cdot \frac{\sigma_{\widehat{X}}}{\sigma_Z} = \frac{Cov(Z,e)}{Cov(\widehat{X},e)} \cdot \frac{\sigma_{\widehat{X}}}{\sigma_Z}
\end{equation}

Conceptually, $\lambda$ measures how $Z$ compares to $\widehat{X}$ in terms of the degree to which each variable violates the exclusion condition, i.e., how much ``better" (or ``worse") $Z$ is than $\widehat{X}$ in this regard. Consequently, $\lambda$ can be used to construct a valid IV, as described in the following Lemma.

\begin{lemma} \label{lem:lambda}
    Let $\widetilde{Z} = \sigma_{\widehat{X}} Z - \lambda \sigma_Z \widehat{X}$, then $\text{Cov}(\widetilde{Z}, u)=0$.
\end{lemma}

Proof of this lemma is included in Appendix \ref{ap:stats_theory}. In a typical measurement error setting, $\lambda$ is unknown because the measurement error component, $e$, is unobservable. However, when the mismeasured variable is generated by a machine learning model, we can empirically estimate $\lambda$ using the labeled data. Specifically, for a pair of individual learners $i \neq j$, where $\widehat{X}^{(i)}$ serves as the endogenous covariate and $\widehat{X}^{(j)}$ serves as the (imperfect) IV, the following Algorithm \ref{algorithm:IV_Transform} first estimates $\lambda$ based on $D_{test}$ and then employs it to transform the IV into a valid instrument.

\begin{algorithm}[tbhp]

\caption{Instrumental Variable Transformation Procedure}
\label{algorithm:IV_Transform}
\KwData{Individual learners $i \neq j$, $D_{test}$, and $D_{unlabel}$}
\tcp{Estimate $\lambda$ from testing data}
Deploy individual learners on $D_{test}$ to get predictions $\widehat{X}_{test}^{(i)}$, $\widehat{X}_{test}^{(j)}$\;
Set $\widehat{X} \gets \widehat{X}_{test}^{(i)}$ and $Z \gets \widehat{X}_{test}^{(j)}$\;
Compute prediction error $e \gets \widehat{X} - X_{test}$\;
Compute $\widehat{\lambda} \gets \frac{Cov(Z,e)}{Cov(\widehat{X},e)} \cdot \frac{\sigma_{\widehat{X}}}{\sigma_Z}$\;
\tcp{Transform IV on unlabeled data}
Deploy individual learners on $D_{unlabel}$ to get predictions $\widehat{X}_{unlabel}^{(i)}$, $\widehat{X}_{unlabel}^{(j)}$\;
Compute $\widetilde{Z}^{(j)} = \sigma_{\widehat{X}} \widehat{X}_{unlabel}^{(j)} - \widehat{\lambda} \sigma_Z \widehat{X}_{unlabel}^{(i)}$\;
\KwOut{$\widehat{X}_{unlabel}^{(i)}$ as the mismeasured covariate and $\widetilde{Z}^{(j)}$ as transformed IV.}
\end{algorithm}

Ideally, to transform the IV on $D_{unlabel}$ (lines 5-6 in Algorithm \ref{algorithm:IV_Transform}), one would need $\widehat{\lambda}_{unlabel}$ defined based on predicted and ground truth values on $D_{unlabel}$. This quantity cannot be estimated in practice (because the ground truth values of $X$ on $D_{unlabel}$ are not observable), which is why we rely on $\widehat{\lambda}$ estimated on $D_{test}$ for IV transformation. Importantly, we establish in Appendix \ref{ap:stats_theory} (specifically, in Lemma \ref{lem:lambda_convergence}) that $\widehat{\lambda}$ estimated on $D_{test}$ tends to (the non-computable) $\widehat{\lambda}_{unlabel}$ (under mild regularity conditions). Therefore, $\widehat{\lambda}$ estimated on a finite (sufficiently large) $D_{test}$ should still provide a reliable approximation of $\widehat{\lambda}_{unlabel}$ to enable IV transformation. 

Taking $\widehat{X}^{(i)}$ as the mismeasured covariate, we apply Algorithm \ref{algorithm:IV_Transform} to each $\widehat{X}^{(j)}$ where $j \in \{1, \ldots, M\} \setminus i$, which produces $M-1$ transformed IVs denoted as $\widetilde{Z}^{(j)}$. Following this procedure will often yield numerous asymptotically valid instruments for the IV estimation, but it may be sub-optimal to employ all of them in this manner for several reasons. First, some IVs may only be weakly correlated with the mismeasured variable, and having weak instruments is known to produce inconsistent estimates \citep{bound1995problems,stock2002testing,andrews2019weak}. Second, with a finite dataset, it is possible that the value of $\widehat{\lambda}$ estimated from $D_{test}$ may not be sufficiently close to (the uncomputable) $\widehat{\lambda}_{unlabel}$ for some pairs of individual learners $i \neq j$. As a result, although the transformed IVs likely improve upon the original (non-transformed) ones in terms of exclusion, they still may not be perfectly excluded. With these ``almost valid" instruments, it becomes even more important to select strong IVs to use in estimation \citep{murray2006avoiding}. Finally, having too many instruments can overfit the endogenous covariate and fail to isolate the exogenous variations \citep{roodman2009note}. Because of these reasons, we next carry out an ``IV selection" step where a subset of strong IVs is selected from the pool of transformed IVs. We specifically consider three different approaches for IV selection, all of which are carried out on $D_{unlabel}$ because this step relies on the transformed IVs $\widetilde{Z}^{(j)}$:
\begin{itemize}
    \item \textbf{Top-$n$}: select the $n$ transformed IVs having the strongest correlation with the mismeasured covariate $\widehat{X}^{(i)}$. This serves as a simple heuristic.
    \item \textbf{PCA}: apply PCA on the transformed IVs $\{ \widetilde{Z}^{(j)} \}_{j \in \{1, \ldots, M\} \setminus i}$, then select the top $n$ components to use as ``condensed" IVs for estimation. This follows the ``factorized IV" approach of \cite{mehrhoff2009solution}. Each principal component, being a linear combination of the transformed IVs, also satisfies the exclusion restriction by definition. Moreover, the top principal components that capture a majority of variations of the transformed IVs are also likely to be strongly correlated with the endogenous covariate, making them good candidates.
    \item \textbf{LASSO}: employ the LASSO-based selection method proposed by \cite{belloni2012sparse}. Roughly speaking, we run a LASSO regression $\widehat{X}^{(i)} \sim \sum_{j \in \{1, \ldots, M\} \setminus i} \gamma_j \widetilde{Z}^{(j)}$ and retain the transformed IVs that exhibit non-zero coefficients. The objective function of this LASSO regression is $\min_{\gamma_j} \left\{\frac{1}{|D_{unlabel}|} \left(\widehat{X}^{(i)} - \sum_{j \in \{1, \ldots, M\} \setminus i} \gamma_j \tilde{Z}^{(j)} \right)^2 + \delta \left(\sum_{j \in \{1, \ldots, M\} \setminus i} |\gamma_j| \right) \right\}$, where $\delta$ is the penalty weight. We adopt the approach of \cite{belloni2012sparse} to determine $\delta$ (see Appendix A of their work for more technical details). \cite{belloni2012sparse} show that this approach will yield a subset of strong IVs that, when employed in an IV regression, produces consistent estimates. 
\end{itemize}

\subsection{EnsembleIV Algorithm} \label{theory_ensembleIV}
Combining the IV transformation and selection procedures, we now describe the EnsembleIV estimation procedure in Algorithm \ref{algorithm:EnsembleIV}. Note that we do not need to specify which technique is used to build the ``ensemble model" because EnsembleIV is generally applicable to ensembles of machine learning models -- this includes common ensemble learning techniques such as random forest, gradient boosting, and stacking, as well as ensembles of different individual learners (e.g., an ensemble of neural networks with varying architectures or hyperparameters).\footnote{In boosting-based ensemble techniques (such as XGBoost or LightGBM), $\widehat{X}^{(i)}$ in Algorithm \ref{algorithm:EnsembleIV} should be the $i$-th \textit{cumulative} learner, as will be discussed in Section \ref{sec:evaluation_boosting}.}

\begin{algorithm}[tbhp]

\caption{EnsembleIV}
\label{algorithm:EnsembleIV}
\KwData{$D_{train}$, $D_{test}$, and $D_{unlabel}$}
Train an ensemble model with $M$ individual learners on $D_{train}$\;
\ForEach{$i \in \{1, \ldots M\}$}{
    Designate $\widehat{X}^{(i)}$ as mismeasured covariate\;
    \tcp{IV Transformation using $D_{test}$ and $D_{unlabel}$}
    Transform each $\widehat{X}^{(j)}, (j \neq i)$ using Algorithm \ref{algorithm:IV_Transform} to obtain $\widetilde{Z}^{(j)}$\;
    \tcp{IV Selection using $D_{unlabel}$}
    Select a subset of strong IVs based on the top-$n$, PCA, or LASSO approach\;
    \tcp{IV Estimation using $D_{unlabel}$}
    Estimate IV regression with the selected $\widetilde{Z}^{(j)}$\;
    Store estimates $\boldsymbol{\widehat{\beta}}_{IV}^{(i)}$\;
}
\KwOut{Average IV estimates $\boldsymbol{\widehat{\beta}}_{IV} = \frac{1}{M} \sum_i \boldsymbol{\widehat{\beta}}_{IV}^{(i)}$.}
\end{algorithm}

The ``IV Estimation" step amounts to any standard IV regression approach established in the econometrics literature, e.g., two-stage least-squares \citep[2SLS,][]{greene2003econometric,wooldridge2002} for linear models and two-stage residual-inclusion \citep[2SRI, also referred to as the control function approach,][]{terza2008two,angrist2008mostly,wan2018general} for generalized linear models (GLMs). Notably, instead of relying on the IV estimates associated with a single individual learner (and its instruments), we average over the IV estimates obtained from each individual learner. Doing so can produce more precise estimations.\footnote{For example, in the case where different $\boldsymbol{\widehat{\beta}}_{IV}^{(i)}$ are independent of each other, averaging can reduce the standard errors by a factor of $\sqrt{M}$. Even if different $\boldsymbol{\widehat{\beta}}_{IV}^{(i)}$ are correlated, variance reduction will still occur, though it will be impaired to the degree correlation exists.}

Under a few mild regularity assumptions, we prove that the EnsembleIV estimator is consistent and asymptotically normal (see Theorem 1 in Appendix \ref{ap:stats_theory}). That said, we also bootstrap our entire procedure to estimate standard errors, given the bootstrap often yields an approximation sampling distribution of an estimator that is more accurate in finite samples than first-order asymptotic approximations~\citep{hall1992bootstrap, horowitz2019bootstrap,davidson2006boostrap}, including in the context of IV regression coefficients~\citep{davidson2008bootstrap}. In particular, we draw from each data partition $\{D_{train}, D_{test}, D_{unlabel}\}$ separately, with replacement, and carry out Algorithm \ref{algorithm:EnsembleIV} on each bootstrapped sample. This represents an end-to-end approach to approximate the uncertainty in EnsembleIV estimates and to support inference. 

\subsection{Enhancing EnsembleIV with Cross-Fitting} \label{theory_crossfitting}
When applying EnsembleIV in practice, the testing partition ($D_{test}$) plays a uniquely important role. It first serves as a dataset to evaluate the out-of-sample performance of the machine learning model trained on $D_{train}$ (enabling model fine-tuning and selection). In addition, it is used for the estimation of $\lambda$ and IV transformation. Given a fixed $D_{label}$, the common practice in building machine learning models is to have a relatively small $D_{test}$ as compared to $D_{train}$. This, however, may lead to noisy $\lambda$ estimates and jeopardize the validity of transformed IVs. At the same time, while having a large $D_{test}$ generally improves $\lambda$ estimates, this results in a small $D_{train}$ partition, which generally produces a poorer machine learning model and greater measurement error.

To address this conundrum, we implement a ``cross-fitting" strategy to enhance EnsembleIV's performance given a relatively small $D_{test}$. The idea of cross-fitting is to use part of the available data to estimate certain nuisance variables (i.e., variables that are not of interest themselves but are needed to estimate the target parameters, such as $\lambda$ in our case), use the rest of the data for target parameter estimation, and then swap the role of each partition and repeat. Finally, we average the resulting estimates of the target parameter to obtain our final estimates. This process allows one to leverage \textit{all} available data for estimation (which improves precision) while continuing to employ different samples of data for nuisance variable and target parameter estimation (thereby avoiding over-fitting). This strategy has proven successful in other contexts, most notably the Double Machine Learning estimator of \cite{chernozhukov2018double}. In Algorithm \ref{algorithm:EnsembleIV_cf}, we integrate cross-fitting with EnsembleIV as part of a sample-splitting procedure.

\begin{algorithm}[tbhp]

\caption{EnsembleIV with Cross-Fitting}
\label{algorithm:EnsembleIV_cf}
\KwData{$D_{label}$ and $K$-fold random partitions $I_1, \ldots, I_k$}
\ForEach{$k \in \{1, \ldots K\}$}{
    Construct $D_{train}$ as all data not in fold $k$ and $D_{test}$ as all data in fold $k$\;
    Run Algorithm \ref{algorithm:EnsembleIV} to obtain $\widehat{\beta}_{IV, k}$\;
}
\KwOut{EnsembleIV estimates with cross-fitting: $\boldsymbol{\widehat{\beta}}_{IV}^{CF} = \frac{1}{K} \sum_k \boldsymbol{\widehat{\beta}}_{IV, k}$.}
\end{algorithm}

Finally, we provide a remark on EnsembleIV's computational complexity. When using EnsembleIV, the most time-consuming aspect will typically relate to the training of the first-phase machine learning model and its application to both $D_{test}$ and $D_{unlabel}$, to obtain predictions from the individual learners. The complexity of this step is increasing in $M$ (although the complexity may vary depending on the specific ensemble learning technique and its software implementation). However, importantly, this step only needs to be done once (or $K$ times with cross-fitting). Next, during the IV transformation and selection step, the theoretical complexity is $O(M^2)$, because the algorithm iteratively designates each individual learner as the endogenous covariate and performs transformation and selection over the other $M-1$ individual learners. That being said, this step is much less time-consuming because IV transformation/selection mainly involves computing correlations across previously calculated variables. Lastly, the IV estimation step involves running $M$ separate IV regressions (or $KM$ regressions with cross-fitting). This step is, again, quite efficient. Overall, the majority of computational overhead relates to training the first-phase machine learning model, while the part of EnsembleIV that is quadratic in $M$ can be implemented very efficiently.

\section{Empirical Evaluations} \label{sec:evaluation}
EnsembleIV offers a systematic approach to generate, transform, select, and use individual learners within an ensemble machine learning model as instrumental variables to address the measurement error problem. In this section, we carry out comprehensive empirical evaluations to understand the effectiveness of the approach. In the first set of evaluations, we rely on synthetic data and simulations to assess EnsembleIV's correction performance for several widely used regression specifications. In the second set of evaluations, we apply EnsembleIV on a real-world dataset consisting of unstructured textual data to demonstrate how the estimator can be used in combination with deep learning techniques.

\subsection{Simulation Studies with Synthetic Data} \label{sec:evaluation_simulation}
Recall the two-phase approach of combining machine learning with statistical inference. In the first, prediction phase, a supervised learning model is trained using the labeled data that can predict the variable of interest in the unlabeled data. In the second, statistical inference phase, the predicted values are incorporated into a regression model as an independent covariate. In our simulation studies, we leverage public datasets to implement the machine learning phase, and we synthesize data atop those real-world samples to inform our assessments of the statistical inference phase.

We employ two publicly available datasets as the basis for the first-phase prediction, namely the ``Bike Sharing" dataset \citep{fanaee2014event} and the ``Bank Marketing" dataset \citep{moro2014data}. Both datasets are available from the UCI Machine Learning Repository\footnote{Bike Sharing data: \url{https://archive.ics.uci.edu/ml/datasets/bike+sharing+dataset}; Bank Marketing data: \url{https://archive.ics.uci.edu/ml/datasets/bank+marketing}.} and are commonly used as benchmarking datasets in the machine learning literature. The Bike Sharing dataset contains 17,379 bike rental records. From this sample, we use date and timestamps in tandem with weather-related features to predict the logarithm of hourly rental volumes (denoted as $lnCnt$). The Bank Marketing dataset contains 45,211 phone calls conducted as part of a Portuguese bank's telemarketing campaign. From this sample, we use client features (e.g., demographics and financial status) to predict conversion, i.e., subscription to a term deposit account with the bank (denoted as $Deposit$). Notably, the target variable to be predicted is continuous in the Bike Sharing dataset and binary in the Bank Marketing dataset, allowing us to evaluate EnsembleIV's performance under both variable types.

We partition each dataset into $D_{label}$ and $D_{unlabel}$. Later in the cross-fitting process, we further partition $D_{label}$ into 4 folds (i.e., $K=4$), equivalent to using 75\% of labeled data for training the machine learning model and 25\% of labeled data for testing. The following Table \ref{table:simulation_setup} summarizes the key aspects of each dataset used in the machine learning phase. In Appendix \ref{ap:descriptive}, we provide descriptive evidence for the effectiveness of the IV transformation and selection steps, i.e., that they can indeed create (approximately) valid instruments and select strong IVs.

\begin{table}[!tbh]
    \centering 
    \caption{Settings in the Machine Learning Phase} \label{table:simulation_setup}
    \begin{tabular}{c|c|c|c|c}
    \hline
    Dataset  & Target Variable &  $|D_{label}|$  &  $|D_{unlabel}|$ & Number of Folds ($K$) \\
    \hline
    Bike Sharing  &  $lnCnt$  &  3000 &  14379  &  4 \\
    \hline
    Bank Marketing  &  $Deposit$  &  4500  &  40711 &  4 \\
    \hline
    \end{tabular}
\end{table}

Next, we simulate data atop these samples for use in the statistical inference phase. We consider two widely-used regression specifications: a linear regression and a logistic regression. The data generation process of each specification is included as follows:

\begin{equation*}
    Y = 1 + 0.5 MLV + 2 W_1 + W_2 + \varepsilon \tag{Linear Regression}
\end{equation*}
\begin{equation*}
    \ln \frac{\Pr(Y=1)}{\Pr(Y=0)} = 1 + 0.5 MLV + 2 W_1 + W_2 \tag{Logistic Regression}
\end{equation*}

where $MLV$ refers to the machine-learning-generated variable (i.e., the target variable in the machine learning phase, $MLV \in \{lnCnt, Deposit\}$); $W_1 \sim Uniform(-10,10)$ and $W_2 \sim N(0, 10^2)$ are independent covariates (i.e., control variables); $\varepsilon \sim N(0, 2^2)$ represents the independent error term in linear specifications; and $Y$ is the dependent variable.

For each specification, we run three sets of regressions: (a) a ``Biased" regression, using the \textit{aggregated prediction} values produced by the random forest for $MLV$ on $D_{unlabel}$; (b) an ``Unbiased" regression, using the \textit{true} values for $MLV$ on $D_{label}$, and (c) the ``EnsembleIV" regression with cross-fitting. While each IV selection method can improve EnsembleIV's estimation, it is hard to determine \textit{a priori} which method would work the best on a specific dataset; we therefore report EnsembleIV estimates under all three methods (i.e., top-3, PCA with top 3 components, and LASSO). The Biased regression produces estimates that one obtains when combining machine learning with statistical inference absent any bias correction effort. The Unbiased regression, in contrast, produces the most precise estimates one can get using the ground truth values available only in the labeled data. We expect the EnsembleIV estimates to be \textit{less biased} than the Biased estimates (due to the use of IV for correction), and \textit{more precise} than the Unbiased estimates (due to its ability to leverage the much larger unlabeled data for estimation).

To obtain the empirical distributions of coefficient estimates, we repeat each simulation experiment, end-to-end, 100 times (including data partitioning, ensemble model training, regression data synthesis, and statistical estimation). We report the mean and standard deviation of each coefficient estimate across 100 simulation runs as the point estimate and standard error, respectively. Further, to quantify the degree of estimation bias, as well as the effectiveness of correction, we report the Estimation Mean-Squared Error (MSE) of each regression, defined as
\begin{equation}
\label{eq:MSE}
    MSE(\boldsymbol{\widehat{\beta}}) = \lVert \boldsymbol{\widehat{\beta}} - \boldsymbol{\beta} \rVert_2 + \sum Var(\boldsymbol{\widehat{\beta}})
\end{equation}
where $\boldsymbol{\beta} = (\beta_0, \beta_{MLV}, \beta_{W_1}, \beta_{W_2}) = (1, 0.5, 2, 1)$ denote the true coefficient values. The estimation MSE is computed as the sum of bias and variance over all coefficients estimated in the regression, with smaller estimation MSE intuitively corresponding to `better' estimates, i.e., estimates that are statistically closer to the true coefficients. We report the simulation results on the Bike Sharing data in Table \ref{table:simulation_main_bike} and those on the Bank Marketing data in Table \ref{table:simulation_main_bank}.

\begin{table}[!tbh]
    \centering 
    \caption{Evaluation Results on Bike Sharing Data}  \label{table:simulation_main_bike}
    \footnotesize \begin{tabular}{c c | c c c c c || c c c c c}
    \hline
    &   & \multicolumn{5}{c||}{Linear Second Phase Regression} & \multicolumn{5}{c}{Logistic Second Phase Regression} \\
    \hline
    & True & Biased & Unbiased & Ens.IV & Ens.IV & Ens.IV & Biased & Unbiased & Ens.IV & Ens.IV & Ens.IV \\ 
    &  &  &  & (Top-3) & (PCA) & (LASSO) &  &  & (Top-3) & (PCA) & (LASSO) \\  
    \hline
    $\beta_0$       & 1.0 & 0.756 & 0.999 & 1.026 & 1.015 & 1.058 & 0.711 & 0.954 & 0.980 & 0.974 & 1.014 \\ 
                    &     & (0.070) & (0.111) & (0.063) & (0.063) & (0.062) & (0.201) & (0.405) & (0.180) & (0.179) & (0.176) \\ 
    $\beta_{MLV}$   & 0.5 & 0.553 & 0.500 & 0.494 & 0.496 & 0.487 & 0.552 & 0.531 & 0.492 & 0.495 & 0.485 \\ 
                    &     & (0.014) & (0.023) & (0.013) & (0.013) & (0.013) & (0.045) & (0.095) & (0.039) & (0.039) & (0.038) \\ 
    $\beta_{W_1}$   & 2.0 & 2.000 & 2.000 & 2.000 & 2.000 & 2.000 & 1.976 & 2.057 & 1.974 & 1.979 & 1.976 \\ 
                    &     & (0.003) & (0.007) & (0.003) & (0.003) & (0.003) & (0.054) & (0.142) & (0.054) & (0.055) & (0.055) \\ 
    $\beta_{W_2}$   & 1.0 & 1.000 & 1.000 & 1.000 & 1.000 & 1.000 & 0.987 & 1.029 & 0.987 & 0.989 & 0.988 \\ 
                    &     & (0.002) & (0.004) & (0.002) & (0.002) & (0.002) & (0.027) & (0.072) & (0.027) & (0.027) & (0.027) \\
    \hline
    Estimation MSE             &     & 0.067 & 0.013 & 0.005 & 0.004 & 0.008 & 0.133 & 0.206 & 0.039 & 0.039 & 0.037 \\
    \hline
    \end{tabular}
\end{table}

\begin{table}[!tbh]
    \centering 
    \caption{Evaluation Results on Bank Marketing Data}  \label{table:simulation_main_bank}
    \footnotesize \begin{tabular}{c c | c c c c c || c c c c c}
    \hline
    &   & \multicolumn{5}{c||}{Linear Second Phase Regression} & \multicolumn{5}{c}{Logistic Second Phase Regression} \\
    \hline
    & True & Biased & Unbiased & Ens.IV & Ens.IV & Ens.IV & Biased & Unbiased & Ens.IV & Ens.IV & Ens.IV \\ 
    &  &  &  & (Top-3) & (PCA) & (LASSO) &  &  & (Top-3) & (PCA) & (LASSO) \\  
    \hline
    $\beta_0$       & 1.0 & 1.044 & 1.002 & 1.008 & 1.006 & 1.017 & 1.041 & 0.980 & 1.003 & 1.002 & 1.012 \\ 
                    &     & (0.011) & (0.028) & (0.013) & (0.013) & (0.012) & (0.036) & (0.102) & (0.041) & (0.041) & (0.040) \\ 
    $\beta_{MLV}$   & 0.5 & 0.280 & 0.503 & 0.427 & 0.442 & 0.353 & 0.292 & 0.505 & 0.443 & 0.458 & 0.367 \\ 
                    &     & (0.043) & (0.096) & (0.049) & (0.050) & (0.040) & (0.140) & (0.291) & (0.166) & (0.169) & (0.136) \\ 
    $\beta_{W_1}$   & 2.0 & 2.000 & 2.000 & 2.000 & 2.000 & 2.000 & 1.999 & 1.998 & 2.001 & 2.001 & 2.000 \\ 
                    &     & (0.002) & (0.005) & (0.002) & (0.002) & (0.002) & (0.029) & (0.094) & (0.029) & (0.029) & (0.029) \\ 
    $\beta_{W_2}$   & 1.0 & 1.000 & 0.999 & 1.000 & 1.000 & 1.000 & 1.000 & 0.998 & 1.000 & 1.001 & 1.000 \\ 
                    &     & (0.001) & (0.003) & (0.001) & (0.001) & (0.001) & (0.015) & (0.047) & (0.015) & (0.015) & (0.015) \\
    \hline
    Estimation MSE             &     & 0.052 & 0.010 & 0.008 & 0.006 & 0.024 & 0.067 & 0.106 & 0.034 & 0.033 & 0.039 \\
    \hline
    \end{tabular}
\end{table}

We observe that EnsembleIV is consistently able to reduce estimation bias (compared to the Biased estimates) and improve estimation precision (compared to the Unbiased estimates). Across all specifications on both datasets, EnsembleIV produces (i) point estimates that are closer to the true coefficient values than the Biased estimates, and (ii) standard errors that are smaller than those of the Unbiased estimates (these precision gains are driven by the much larger $D_{unlabel}$). EnsembleIV's advantages are also reflected in the estimation MSE metric -- EnsembleIV estimates have lower estimation MSE than both the Biased and Unbiased estimates (with a single exception -- the LASSO-selected IVs and linear second phase regression on the Bank Marketing dataset -- though we still observe bias reduction). Further, IV selection based on Top-3 and PCA largely results in similar EnsembleIV estimates, whereas the LASSO selection approach tends to produce somewhat worse point estimates, yet lower standard errors. The above simulation studies demonstrate EnsembleIV's utility to mitigate estimation biases caused by the measurement error problem, while also leveraging the larger $D_{unlabel}$ to improve estimation precision.

\subsection{Apply EnsembleIV with Boosting} \label{sec:evaluation_boosting}
An important advantage of EnsembleIV is that it can be applied with different common ensemble learning techniques. Owing to the IV transformation step, valid IVs can be created even if the resulting individual learners do not have inherently low error correlations. In this section, we demonstrate how EnsembleIV can be used with gradient boosting and evaluate its performance using the two synthetic datasets.

We choose gradient boosting here because it represents a starkly different type of ensemble learning technique than random forest. Under gradient boosting, a collection of $M$ boosting trees are trained in a \textit{sequential} manner, which we denote with a tuple $(1, \ldots, M)$. Each boosting tree $i > 1$ aims to predict the errors (i.e., residuals) of the earlier boosting trees $1 \leq j < i$. To apply EnsembleIV with gradient boosting, we first use the original $M$ boosting trees to construct $M$ ``cumulative" learners, where the $i$-th cumulative learner is the aggregation of all boosting trees $j \in (1, \ldots, i)$. We use the predictions from these $M$ cumulative learners as the endogenous variables and their candidate instruments. Intuitively, different cumulative learners are all somewhat predictive of the target ground truth, thereby supporting the relevance condition of IVs. However, unlike in a random forest, there is little reason to believe that the prediction errors of cumulative learner $i$ are only weakly correlated with the predictions of cumulative learner $j$, because the two cumulative learners have a non-trivial overlap in their constituent boosting trees. The likely violation of the exclusion condition makes this a challenging and meaningful test of EnsembleIV's ability to create valid IVs from data.

We apply EnsembleIV with XGBoost \citep{chen2016xgboost}, a popular gradient boosting algorithm, on both the Bike Sharing dataset and the Bank Marketing dataset with a linear and logit second-phase regression, respectively. The simulation setups are the same as before, and the results are reported in Tables \ref{table:simulation_boosting_bike} and \ref{table:simulation_boosting_bank} for the two datasets, respectively.

\begin{table}[!tbh]
    \centering 
    \caption{Evaluation Results on Bike Sharing Data with XGBoost} \label{table:simulation_boosting_bike} 
    \footnotesize \begin{tabular}{c c | c c c c c || c c c c c}
    \hline
    &   & \multicolumn{5}{c||}{Linear Second Phase Regression} & \multicolumn{5}{c}{Logistic Second Phase Regression} \\
    \hline
    & True & Biased & Unbiased & Ens.IV & Ens.IV & Ens.IV & Biased & Unbiased & Ens.IV & Ens.IV & Ens.IV \\ 
    &  &  &  & (Top-3) & (PCA) & (LASSO) &  &  & (Top-3) & (PCA) & (LASSO) \\  
    \hline
    $\beta_0$       & 1.0 & 1.029 & 0.999 & 1.060 & 1.060 & 1.059 & 1.013 & 1.006 & 1.043 & 1.043 & 1.044 \\ 
                    &     & (0.058) & (0.111) & (0.060) & (0.060) & (0.059) & (0.165) & (0.439) & (0.169) & (0.167) & (0.167) \\ 
    $\beta_{MLV}$   & 0.5 & 0.494 & 0.500 & 0.497 & 0.497 & 0.498 & 0.495 & 0.513 & 0.500 & 0.500 & 0.500 \\ 
                    &     & (0.012) & (0.023) & (0.013) & (0.013) & (0.012) & (0.039) & (0.086) & (0.041) & (0.040) & (0.040) \\ 
    $\beta_{W_1}$   & 2.0 & 2.000 & 2.000 & 2.000 & 2.000 & 2.000 & 1.986 & 2.020 & 1.987 & 1.987 & 1.987 \\ 
                    &     & (0.003) & (0.007) & (0.003) & (0.003) & (0.003) & (0.052) & (0.148) & (0.052) & (0.052) & (0.052) \\ 
    $\beta_{W_2}$   & 1.0 & 1.000 & 1.000 & 1.000 & 1.000 & 1.000 & 0.994 & 1.011 & 0.994 & 0.994 & 0.994 \\ 
                    &     & (0.002) & (0.004) & (0.002) & (0.002) & (0.002) & (0.026) & (0.075) & (0.026) & (0.026) & (0.026) \\
    \hline
    Estimation MSE             &     & 0.004 & 0.013 & 0.007 & 0.007 & 0.007 & 0.033 & 0.228 & 0.036 & 0.035 & 0.035 \\
    \hline
    \end{tabular}
\end{table}

\begin{table}[!tbh]
    \centering 
    \caption{Evaluation Results on Bank Marketing Data with XGBoost} \label{table:simulation_boosting_bank} 
    \footnotesize \begin{tabular}{c c | c c c c c || c c c c c}
    \hline
    &   & \multicolumn{5}{c||}{Linear Second Phase Regression} & \multicolumn{5}{c}{Logistic Second Phase Regression} \\
    \hline
    & True & Biased & Unbiased & Ens.IV & Ens.IV & Ens.IV & Biased & Unbiased & Ens.IV & Ens.IV & Ens.IV \\ 
    &  &  &  & (Top-3) & (PCA) & (LASSO) &  &  & (Top-3) & (PCA) & (LASSO) \\  
    \hline
    $\beta_0$       & 1.0 & 1.037 & 1.002 & 1.002 & 1.001 & 1.005 & 1.032 & 0.979 & 0.997 & 0.996 & 0.999 \\ 
                    &     & (0.012) & (0.028) & (0.013) & (0.013) & (0.013) & (0.036) & (0.103) & (0.042) & (0.042) & (0.042) \\ 
    $\beta_{MLV}$   & 0.5 & 0.253 & 0.503 & 0.475 & 0.482 & 0.443 & 0.270 & 0.513 & 0.499 & 0.508 & 0.475 \\ 
                    &     & (0.029) & (0.096) & (0.056) & (0.059) & (0.047) & (0.104) & (0.286) & (0.180) & (0.185) & (0.166) \\ 
    $\beta_{W_1}$   & 2.0 & 2.000 & 2.000 & 2.000 & 2.000 & 2.000 & 2.000 & 1.998 & 2.001 & 2.001 & 2.001 \\ 
                    &     & (0.002) & (0.005) & (0.002) & (0.002) & (0.002) & (0.029) & (0.095) & (0.029) & (0.029) & (0.031) \\ 
    $\beta_{W_2}$   & 1.0 & 1.000 & 0.999 & 1.000 & 1.000 & 1.000 & 1.000 & 0.997 & 1.001 & 1.001 & 1.001 \\ 
                    &     & (0.001) & (0.003) & (0.001) & (0.001) & (0.001) & (0.015) & (0.047) & (0.015) & (0.015) & (0.016) \\
    \hline
    Estimation MSE             &     & 0.063 & 0.010 & 0.004 & 0.004 & 0.006 & 0.067 & 0.104 & 0.035 & 0.037 & 0.031 \\
    \hline
    \end{tabular}
\end{table}

On the Bike Sharing dataset, the XGBoost model achieves a prediction RMSE of 0.55 (evaluated on $D_{test}$). This is quite low considering that the target ground truth ($lnCnt$) ranges from 0 to 6.88 with a standard deviation of 1.48. Accordingly, the relatively small measurement error in the Biased regression does not result in substantial estimation biases (especially under the linear regression specification). In this case, we observe that EnsembleIV estimates are still very close to the true coefficient values though with smaller standard errors and estimation MSEs than the Unbiased estimates. In other words, EnsembleIV does not introduce any additional biases in cases where measurement error is small to begin with. On the Bank Marketing dataset, directly using the XGBoost model's predictions in the Biased regression has resulted in over 50\% underestimation of $\beta_{MLV}$. Importantly, EnsembleIV is again able to mitigate the estimation biases and maintain smaller standard errors and estimation MSEs than the Unbiased estimates. All three IV selection approaches produce similar EnsembleIV estimates in terms of the estimation MSE scores. On the Bank Marketing dataset, we again observe that the LASSO selection approach leads to slightly worse point estimates yet smaller standard errors than the other two selection approaches.\footnote{Based on the simulation results in Tables \ref{table:simulation_main_bike}-\ref{table:simulation_boosting_bank}, we note that the performance of the different IV selection methods in \textit{finite samples} is likely affected by idiosyncrasies in data, machine learning techniques, and second phase regression specifications. In practice, instead of choosing one selection method up front, researchers could repeat estimation with different methods to establish robustness.}

\section{Benchmarking and Robustness Studies} \label{sec:benchmark}
While the previous section demonstrates EnsembleIV's bias correction effectiveness on both synthetic and real-world datasets, in this section we focus on benchmarking EnsembleIV against alternative bias correction approaches and evaluating EnsembleIV's performance robustness. For simplicity, we present EnsembleIV results with PCA-based IV selection in this section, although the other selection methods yield qualitatively similar findings.

\subsection{Benchmarking with ForestIV} \label{sec:benchmark_forestiv}
Using the same simulation setup as before, we compare the correction performance between EnsembleIV (with 4-fold cross-fitting) and its ``closest kin", namely ForestIV \citep[see][for implementation details of the ForestIV algorithm]{yang2022achieving}. We note that the original design of ForestIV does not have a cross-fitting component. For a more comprehensive comparison, we therefore report ForestIV estimates both with and without 4-fold cross-fitting. The results are presented in Tables \ref{table:benchmark_forestiv_bike} and \ref{table:benchmark_forestiv_bank}, for the two datasets, respectively.

\begin{table}[!tbh]
    \centering 
    \caption{Benchmarking with ForestIV on Bike Sharing Data} \label{table:benchmark_forestiv_bike} 
    \begin{tabular}{c c | c c c || c c c}
    \hline
    &   & \multicolumn{3}{c||}{Linear Second Phase Regression} & \multicolumn{3}{c}{Logistic Second Phase Regression} \\
    \hline
    & True & Ens.IV & ForestIV & ForestIV & Ens.IV & ForestIV & ForestIV \\ 
    &  &  & (w/o CF) & (with CF) &  & (w/o CF) & (with CF) \\  
    \hline
    $\beta_0$       & 1.0 & 1.015 & 0.939 & 0.938 & 0.974 & 0.847 & 0.850 \\ 
                    &     & (0.063) & (0.074) & (0.073) & (0.179) & (0.206) & (0.206) \\ 
    $\beta_{MLV}$   & 0.5 & 0.496 & 0.514 & 0.515 & 0.495 & 0.530 & 0.530 \\ 
                    &     & (0.013) & (0.016) & (0.016) & (0.039) & (0.047) & (0.046) \\ 
    $\beta_{W_1}$   & 2.0 & 2.000 & 2.000 & 2.000 & 1.979 & 1.989 & 1.989 \\ 
                    &     & (0.003) & (0.003) & (0.003) & (0.055) & (0.056) & (0.056) \\ 
    $\beta_{W_2}$   & 1.0 & 1.000 & 1.000 & 1.000 & 0.989 & 0.995 & 0.995 \\ 
                    &     & (0.002) & (0.002) & (0.002) & (0.027) & (0.029) & (0.029) \\
    \hline
    Estimation MSE             &     & 0.004 & 0.010 & 0.010 & 0.039 & 0.073 & 0.072 \\
    \hline
    \end{tabular}
\end{table}

\begin{table}[!tbh]
    \centering 
    \caption{Benchmarking with ForestIV on Bank Marketing Data} \label{table:benchmark_forestiv_bank} 
    \begin{tabular}{c c | c c c || c c c}
    \hline
    &   & \multicolumn{3}{c||}{Linear Second Phase Regression} & \multicolumn{3}{c}{Logistic Second Phase Regression} \\
    \hline
    & True & Ens.IV & ForestIV & ForestIV & Ens.IV & ForestIV & ForestIV \\ 
    &  &  & (w/o CF) & (with CF) &  & (w/o CF) & (with CF) \\  
    \hline
    $\beta_0$       & 1.0 & 1.006 & 0.994 & 0.994 & 1.002 & 0.994 & 0.993 \\ 
                    &     & (0.013) & (0.012) & (0.012) & (0.041) & (0.042) & (0.043) \\ 
    $\beta_{MLV}$   & 0.5 & 0.442 & 0.539 & 0.538 & 0.458 & 0.550 & 0.551 \\ 
                    &     & (0.050) & (0.066) & (0.067) & (0.169) & (0.175) & (0.172) \\ 
    $\beta_{W_1}$   & 2.0 & 2.000 & 2.000 & 2.000 & 2.001 & 1.994 & 1.994 \\ 
                    &     & (0.002) & (0.002) & (0.002) & (0.029) & (0.034) & (0.034) \\ 
    $\beta_{W_2}$   & 1.0 & 1.000 & 1.000 & 1.000 & 1.001 & 0.997 & 0.997 \\ 
                    &     & (0.001) & (0.001) & (0.001) & (0.015) & (0.017) & (0.017) \\
    \hline
    Estimation MSE             &     & 0.006 & 0.006 & 0.006 & 0.033 & 0.037 & 0.036 \\ 
    \hline
    \end{tabular}
\end{table}

We find that EnsembleIV estimates have lower standard errors than ForestIV estimates whether or not we implement ForestIV with cross-fitting. This offers empirical support to EnsembleIV's advantage in estimation efficiency. Point estimates from EnsembleIV are also closer to the true coefficient value than those from ForestIV, except in the case of a linear second-phase regression on Bank Marketing data. Overall, this set of benchmarking analyses suggests that EnsembleIV represents a more efficient estimator than ForestIV (even after the enhancement of cross-fitting). EnsembleIV's efficiency benefits can be attributed to its design. ForestIV identifies the single ``best" combination of individual learners that would yield estimates exhibiting the smallest deviation from the unbiased estimates obtained on $D_{label}$, whereas EnsembleIV averages over all combinations. Accordingly, in a given dataset, EnsembleIV will generally be more efficient than ForestIV.

Importantly, we note that EnsembleIV's advantage over ForestIV also goes beyond estimation efficiency. First, the validity of ForestIV depends on the specific theoretical properties of a single machine learning technique, the random forest algorithm. Further, that validity has only been formally established for scenarios involving continuous machine-learning-generated variables \cite{yang2022achieving}. In contrast, EnsembleIV can be applied with ensembles of machine learning models -- this includes common ensemble learning techniques such as random forest, gradient boosting, and stacking, as well as ensembles of different individual learners (e.g., an ensemble of neural networks with varying architectures or hyperparameters). EnsembleIV is also provably valid for both binary and continuous machine-learning-generated variables (see EnsembleIV's asymptotic properties, shown in Appendix \ref{ap:stats_theory}, which do not depend on the type of machine-learning-generated variables). Second, whereas ForestIV is passive, relying on the chance discovery of individual learners from a trained random forest that happens to yield predictions meeting the exclusion condition, EnsembleIV is active, purposefully \textit{transforming} all candidate instruments to achieve compliance with the exclusion condition. For this reason, EnsembleIV has a more solid theoretical grounding and it provides the ability to average over multiple IV estimates to improve estimation efficiency.\footnote{We note that the manuscript introducing ForestIV \citep{yang2022achieving} also experimented with the idea of using subsets of trees as endogenous covariates and IVs \citep[see Appendix C of][]{yang2022achieving}, but the authors found it to have limited effectiveness. We repeat the same exploration and again find it to result in generally worse correction performance. See Appendix \ref{ap:subset} for more details.}

\subsection{Benchmarking with Other Correction Approaches} \label{sec:benchmark_other}
Next, we benchmark EnsembleIV with two other bias correction approaches, respectively Regression Calibration \citep{gleser1992importance,carroll1995measurement} and a GMM estimator proposed by \cite{fong2021machine}, which we refer to as ``FT-GMM" for short. While Regression Calibration is a classic method from the measurement error literature, FT-GMM represents a recently developed method to address the same estimation bias problem that we consider in this paper.\footnote{Another representative bias correction method is Simulation Extrapolation \citep[SIMEX,][]{Stefanski1995,Kuchenhoff2006}. We note that ForestIV has been shown to outperform SIMEX in \cite{yang2022achieving}, and we therefore omit SIMEX as a benchmark here.}

The Regression Calibration approach relies on a subsample of data where both the ground truth values and the predicted values of the machine-learning-generated variable can be observed, i.e., $D_{test}$ in our context. Using this set of data, it learns a \textit{calibration model} by regressing the ground truth values onto the predicted values (controlling for other exogenous covariates). Then, the calibration model is applied to $D_{unlabel}$ to obtain calibrated predictions. Finally, the second-phase regression model is estimated using those calibrated predictions (rather than the original predictions from the first-phase machine learning model). 

The FT-GMM approach starts by following the same steps as Regression Calibration --- using $D_{test}$ (``validation data'' in the authors' notation) to learn a projection of the ground truth values onto the predicted values (although the authors interpreted this step as the first phase of a 2SLS estimation), then apply this projection to $D_{unlabel}$ (``primary data'' in the authors' notation). Then, unlike Regression Calibration, the FT-GMM approach seeks to \textit{combine} the 2SLS estimates on $D_{unlabel}$ with the OLS estimates on $D_{train} \cup D_{test}$ via GMM, concatenating the moment conditions of the two estimators. Doing so potentially allows FT-GMM to achieve higher estimation efficiency. Importantly, however, the proposed FT-GMM approach does not support GLMs, hence we cannot apply it to a scenario involving Logistic regression in the second phase.

Therefore, we apply both benchmark approaches to the Bike Sharing and Bank Marketing datasets in the case of a linear second-phase regression. However, we limit our consideration to the Regression Calibration approach when it comes to a Logistic second-phase regression. In each case, we compare the benchmark methods' correction performance with that of EnsembleIV (with 4-fold cross-fitting). Neither benchmark was initially designed with the cross-fitting component; however, we report their performance both with and without cross-fitting, for the sake of completeness. The results are presented in Tables \ref{table:benchmark_other_bike} and \ref{table:benchmark_other_bank}.

\begin{table}[!tbh]
    \centering 
    \caption{Benchmarking with Regression Calibration and FT-GMM on Bike Sharing Data} \label{table:benchmark_other_bike} 
    \footnotesize \begin{tabular}{c c | c c c c c || c c c}
    \hline
    &   & \multicolumn{5}{c||}{Linear Second Phase Regression} & \multicolumn{3}{c}{Logistic Second Phase Regression} \\
    \hline
    & True & Ens.IV & Reg.Cal & Reg.Cal & FT-GMM & FT-GMM & Ens.IV & Reg.Cal & Reg.Cal \\ 
    &  &  & (w/o CF) & (with CF) & (w/o CF) & (with CF) &  & (w/o CF) & (with CF) \\  
    \hline
    $\beta_0$       & 1.0 & 1.015 & 1.036 & 1.032 & 1.042 & 1.039 & 0.974 & 1.202 & 1.196 \\ 
                    &     & (0.063) & (0.050) & (0.033) & (0.039) & (0.024) & (0.179) & (0.096) & (0.088) \\ 
    $\beta_{MLV}$   & 0.5 & 0.496 & 0.489 & 0.489 & 0.487 & 0.488 & 0.495 & 0.444 & 0.445 \\ 
                    &     & (0.013) & (0.009) & (0.006) & (0.007) & (0.005) & (0.039) & (0.019) & (0.018) \\ 
    $\beta_{W_1}$   & 2.0 & 2.000 & 1.998 & 1.998 & 1.998 & 1.998 & 1.979 & 1.971 & 1.971 \\ 
                    &     & (0.003) & (0.002) & (0.001) & (0.001) & (0.001) & (0.055) & (0.025) & (0.025) \\ 
    $\beta_{W_2}$   & 1.0 & 1.000 & 1.000 & 1.000 & 1.000 & 1.000 & 0.989 & 0.992 & 0.992 \\ 
                    &     & (0.002) & (0.001) & (0.001) & (0.001) & (0.000) & (0.027) & (0.013) & (0.013) \\
    \hline
    Estimation MSE             &     & 0.004 & 0.004 & 0.002 & 0.004 & 0.002 & 0.039 & 0.055 & 0.051 \\
    \hline
    \end{tabular}
\end{table}

\begin{table}[!tbh]
    \centering 
    \caption{Benchmarking with Regression Calibration and FT-GMM on Bank Marketing Data} \label{table:benchmark_other_bank} 
    \footnotesize \begin{tabular}{c c | c c c c c || c c c}
    \hline
    &   & \multicolumn{5}{c||}{Linear Second Phase Regression} & \multicolumn{3}{c}{Logistic Second Phase Regression} \\
    \hline
    & True & Ens.IV & Reg.Cal & Reg.Cal & FT-GMM & FT-GMM & Ens.IV & Reg.Cal & Reg.Cal \\ 
    &  &  & (w/o CF) & (with CF) & (w/o CF) & (with CF) &  & (w/o CF) & (with CF) \\  
    \hline
    $\beta_0$       & 1.0 & 1.006 & 0.997 & 0.997 & 0.999 & 0.999 & 1.002 & 1.029 & 1.031 \\ 
                    &     & (0.013) & (0.010) & (0.008) & (0.007) & (0.006) & (0.041) & (0.029) & (0.022) \\ 
    $\beta_{MLV}$   & 0.5 & 0.442 & 0.491 & 0.490 & 0.476 & 0.475 & 0.458 & 0.467 & 0.447 \\ 
                    &     & (0.050) & (0.079) & (0.060) & (0.061) & (0.055) & (0.169) & (0.218) & (0.161) \\ 
    $\beta_{W_1}$   & 2.0 & 2.000 & 2.001 & 2.001 & 2.001 & 2.001 & 2.001 & 1.987 & 1.987 \\ 
                    &     & (0.002) & (0.001) & (0.001) & (0.001) & (0.000) & (0.029) & (0.011) & (0.011) \\ 
    $\beta_{W_2}$   & 1.0 & 1.000 & 1.002 & 1.002 & 1.002 & 1.002 & 1.001 & 0.990 & 0.990 \\ 
                    &     & (0.001) & (0.001) & (0.000) & (0.000) & (0.000) & (0.015) & (0.006) & (0.006) \\
    \hline
    Estimation MSE             &     & 0.006 & 0.006 & 0.004 & 0.004 & 0.004 & 0.033 & 0.051 & 0.031 \\
    \hline
    \end{tabular}
\end{table}

With a linear second-phase regression, all three methods arrive at fairly similar estimation results on the Bike Sharing data. EnsembleIV produces better point estimates whereas the two benchmarks have smaller standard errors. After applying the cross-fitting enhancement, the two benchmarks can achieve slightly smaller empirical estimation MSE values. In comparison, on the Bank Marketing data, EnsembleIV has smaller standard errors than the benchmarks (even after applying cross-fitting), though the point estimate is not as close to the true value. Under a logistic second-phase regression, EnsembleIV achieves a smaller estimation MSE than Regression Calibration in most cases, except on Bank Marketing data where EnsembleIV has a similar estimation MSE as the cross-fitting-enhanced Regression Calibration. Overall, EnsembleIV performs competitively against the benchmarks and, unlike FT-GMM, naturally accommodates GLM second-phase regressions without the need for regression-specific modifications to the correction procedure.

\subsection{Sensitivity Analyses} \label{sec:sensitivity_performance}
In addition to benchmarking, we also derive a better understanding of EnsembleIV's performance sensitivity with changes to several parameters that researchers have to decide upon when applying the approach:
\begin{itemize}
    \item $|D_{label}|$, the amount of labeled data available;
    \item $M$, the number of individual learners in the first-phase ensemble model;
    \item $\sigma_{\varepsilon}$, the amount of noise in the second-phase regression model, which is affected by how many relevant control variables are included in the regression;
    \item $K$, the number of folds for cross-fitting;
\end{itemize}
For simplicity, we focus on the Bike Sharing data and the linear second-phase regression. We start with the same parameter configuration as our main simulation studies (Table \ref{table:simulation_setup}) and vary one parameter at a time in a series of experiments. Specifically, we vary $|D_{label}| \in \{1000, 2000, 3000, 4000, 5000\}$, $M \in \{50, 100, 150, 200, 250\}$, $\sigma_{\varepsilon} \in \{0.5, 1.0, 2.0, 3.0, 4.0\}$, and $K \in \{2, 3, 4, 5, 6\}$. In the following Figure \ref{fig:Sensitivity_Parameters_bike}, we plot the point estimates of the machine-learning-generated covariate and its empirical 95\% confidence intervals of the Biased, Unbiased, and EnsembleIV estimators as we vary each parameter.

\begin{figure}[!tbh]
    \centering
    \includegraphics[width=\linewidth]{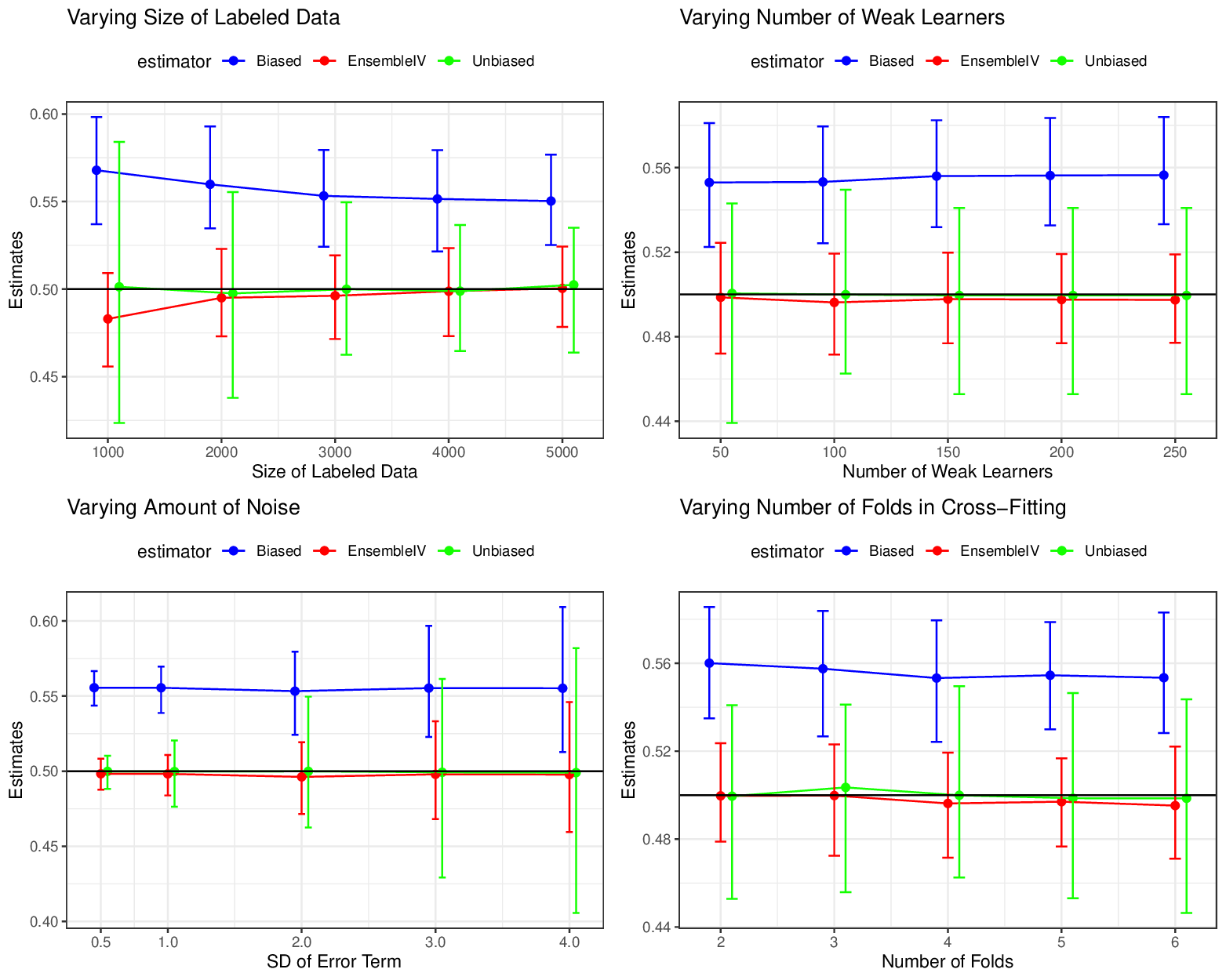}
    \caption{EnsembleIV performance under different parameter configurations (Bike Sharing Data). Vertical bars represent empirical 95\% confidence intervals. Black horizontal lines mark the true coefficient value (0.5).}
    \label{fig:Sensitivity_Parameters_bike}
\end{figure}

We discuss several patterns that can potentially assist researchers in setting these parameters. In the top-left plot, as more labeled data becomes available, the EnsembleIV point estimate converges nicely to the true value of the coefficient. Such a convergence is likely the result of two factors. First, having a larger $|D_{label}|$ means that more training data (i.e., $\frac{K-1}{K}$ of $|D_{label}|$) are available, which can improve the predictive performance of the first-phase ensemble model. Indeed, across the five sizes of $|D_{label}|$, we observe a consistent decline of prediction RMSE values (averaged over 4-fold cross-validation) of 0.68, 0.60, 0.58, 0.55, and 0.52, aligned with better EnsembleIV estimates. Second, increasing $|D_{label}|$ also results in more testing data (i.e., $\frac{1}{K}$ of $|D_{label}|$), which benefits the precision of $\lambda$ estimation and IV transformation. Meanwhile, we note that the confidence interval of the Unbiased estimate expectedly shrinks. The top-right plot shows that EnsembleIV estimates are somewhat insensitive with respect to the number of individual learners. This is likely because, on the Bike Sharing data, varying $M$ from 50 to 250 does not significantly affect the predictive performance.\footnote{The prediction RMSE values (averaged over 4-fold cross-validation) are 0.572, 0.576, 0.577, 0.568, 0.569, respectively.}

The bottom-left plot shows that, as the standard deviation of the regression error term becomes larger, all three estimators have wider confidence intervals, while the point estimates of EnsembleIV and Unbiased regression remain very close to the true coefficient value. This demonstrates the robustness of EnsembleIV to increases in exogenous noise in the second-phase regression model. Hence, efforts to include relevant control variables in the regression, which can reduce the standard deviation of the error term, would benefit EnsembleIV by producing more precise estimates. Finally, the bottom-right plot indicates that EnsembleIV is fairly insensitive with respect to the number of folds used in cross-fitting on the Bike Sharing dataset, suggesting that it is not necessary to use a large number of folds when applying EnsembleIV, given cross-fitting with more folds is computationally more costly.

\begin{figure}[!tbh]
    \centering
    \includegraphics[width=\linewidth]{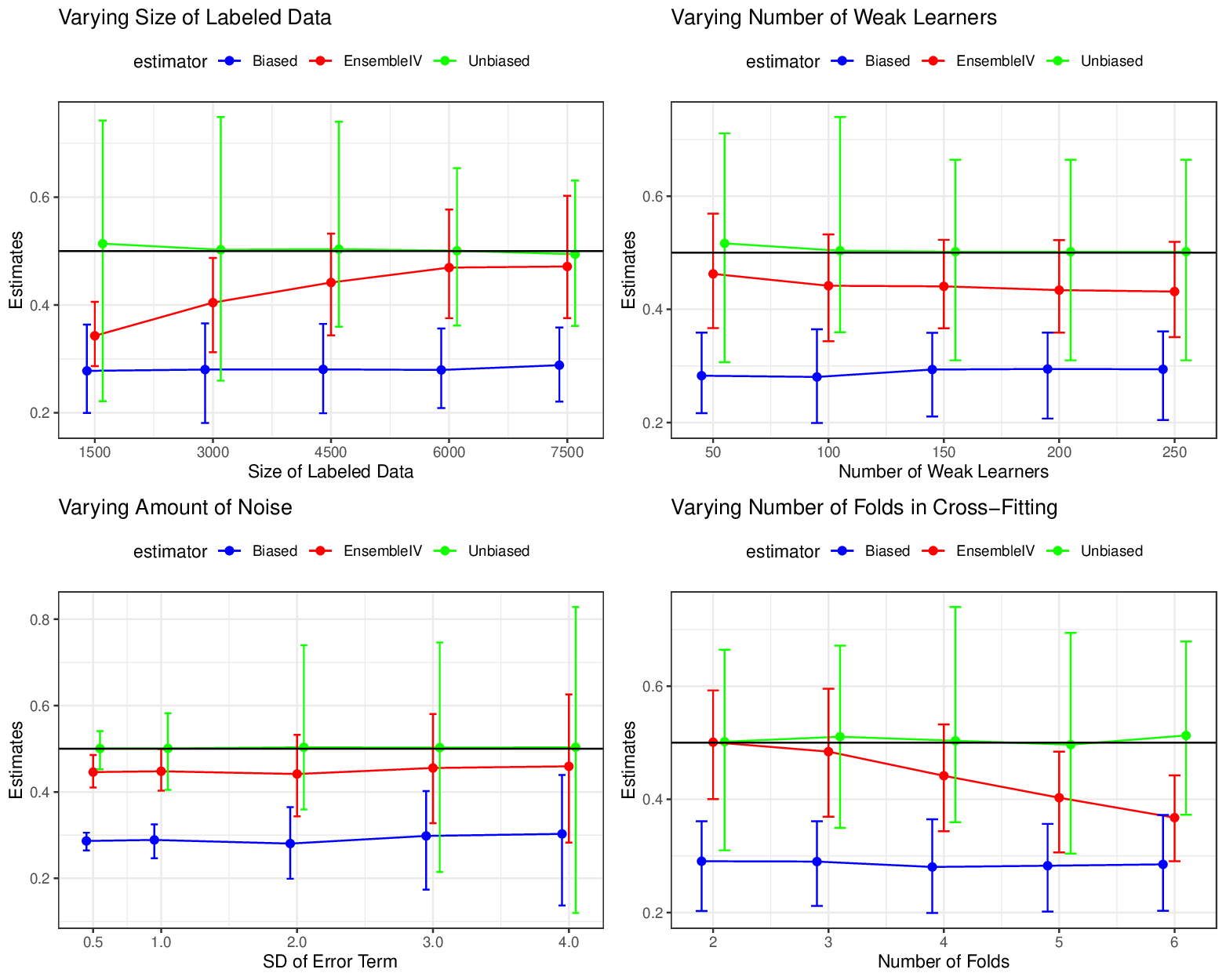}
    \caption{EnsembleIV performance under different parameter configurations (Bank Marketing Data). Vertical bars represent empirical 95\% confidence intervals. Black horizontal lines mark the true coefficient value (0.5).}
    \label{fig:Sensitivity_Parameters_bank}
\end{figure}

We repeat these sensitivity analyses on the Bank Marketing dataset. Because this dataset is larger, we also impose a larger increment in the size of labeled data, i.e., $D_{label} \in \{1500, 3000, 4500, 6000, 7500\}$. Variations of the other parameters are kept the same as above. The results are presented in Figure \ref{fig:Sensitivity_Parameters_bank}.

We observe patterns consistent with those noted above. EnsembleIV again shows a trend of convergence as the size of labeled data increases, given the increased accuracy of the first-phase predictive model and more testing data for IV transformation. As we increase the number of individual learners from 50 to 100, there is a bit more bias in EnsembleIV's point estimates -- this also aligns with the fact that the first-phase random forest model with 50 trees achieved a slightly smaller validation loss than that with 100 trees. Overall, our results when varying $D_{label}$ and $M$ indicate that having a better-performing first-phase model generally improves EnsembleIV's correction performance. Therefore, researchers' efforts to fine-tune the first-phase machine learning models (e.g., via parameter tuning or model selection) are instrumental to EnsembleIV's performance. Notably, in the bottom-right plot, we see that a larger number of folds in cross-fitting hurts EnsembleIV's performance, likely because the size of testing data becomes too small to enable sufficiently good IV transformation. Considering these observations in tandem with our observations on Bike Sharing data, we recommend using a small number of folds in cross-fitting. Doing so simultaneously benefits correction performance and reduces computational overhead.

In addition to the parameters examined above, researchers also need to decide $n$, the number of strong IVs, if Top-$n$ or PCA-based methods are used for IV selection. In the case of Top-$n$, selected IVs are those having the strongest correlation with the endogenous covariate; and in the case of PCA, the top $n$ components (each being a linear combination of original IVs) are also those that account for the most variation in the endogenous covariate. Theoretically, using a small number of strong IVs is preferred over using a large number of potentially weaker IVs because the latter can lead to misleading inferential results \citep{stock2002survey}. Under the same empirical setup as before, we vary the number of top IVs used from 1 to 5. The results, summarized in Table \ref{table:topIV}, show that changes in the number of IVs yield very little differences, across the board. Therefore, a small number of strong IVs is generally sufficient in practice. In terms of \textit{exactly} how many strong IVs one should use, we note that \cite{donald2001choosing} propose an approach based on minimizing an approximated MSE, which can serve as a guideline.

\begin{table}[!tbh]
    \centering 
    \caption{EnsembleIV Estimates with Different Number of Strong IVs (Bike Sharing Data)} \label{table:topIV}
    \footnotesize \begin{tabular}{c c | c c c c c || c c c c c}
    \hline
    &   & \multicolumn{5}{c||}{Top-$n$} & \multicolumn{5}{c}{PCA} \\
    \hline
    & True & 1 & 2 & 3 & 4 & 5 & 1 & 2 & 3 & 4 & 5 \\ 
    \hline
    $\beta_0$       & 1.0 & 1.025 & 1.024 & 1.023 & 1.022 & 1.022 & 1.001 & 1.009 & 1.013 & 1.018 & 1.022 \\ 
                    &     & (0.061) & (0.061) & (0.061) & (0.061) & (0.061) & (0.061) & (0.062) & (0.061) & (0.061) & (0.061) \\ 
    $\beta_{MLV}$   & 0.5 & 0.495 & 0.495 & 0.496 & 0.496 & 0.496 & 0.500 & 0.499 & 0.498 & 0.498 & 0.497 \\ 
                    &     & (0.013) & (0.013) & (0.013) & (0.013) & (0.013) & (0.013) & (0.013) & (0.013) & (0.013) & (0.013) \\ 
    $\beta_{W_1}$   & 2.0 & 2.001 & 2.001 & 2.001 & 2.001 & 2.001 & 2.001 & 2.001 & 2.001 & 2.001 & 2.001 \\ 
                    &     & (0.003) & (0.003) & (0.003) & (0.003) & (0.003) & (0.003) & (0.003) & (0.003) & (0.003) & (0.003) \\ 
    $\beta_{W_2}$   & 1.0 & 1.000 & 1.000 & 1.000 & 1.000 & 1.000 & 1.000 & 1.000 & 1.000 & 1.000 & 1.000 \\ 
                    &     & (0.002) & (0.002) & (0.002) & (0.002) & (0.002) & (0.002) & (0.002) & (0.002) & (0.002) & (0.002) \\
    \hline
    Estimation MSE             &     & 0.005 & 0.004 & 0.004 & 0.004 & 0.004 & 0.004 & 0.004 & 0.004 & 0.004 & 0.004 \\ 
    \hline
    \end{tabular}
\end{table}

Finally, we note that EnsembleIV's correction performance under different IV selection methods (and with different numbers of selected IVs) in \textit{finite samples} is likely affected by idiosyncrasies in data, machine learning techniques, and second-phase regression specifications. Accordingly, in practice, researchers may be best served by repeating estimation with different selection methods to establish robustness.

\section{Field Evaluation: Engagement with User-Generated Content on Facebook} \label{sec:evaluation_facebook}
We now conduct an in-vivo evaluation of EnsembleIV using a real-world dataset collected from Facebook business pages. Business pages on Facebook are public spaces managed by firms to share marketing content (e.g., product promotions), who seek to interact with Facebook users (e.g., answering their questions or complaints). Several prior studies have examined firm- and user-generated content in this context, and their impact on user engagement or firm performance \citep[e.g.,][]{goh2013social,lee2018advertising,yang2019understanding}. For this evaluation, we estimate how the sentiment (positive or negative) of a user-generated post affects content engagement, measured as the number of comments that the post receives from other users. This evaluation serves a few purposes. First, it represents a realistic empirical context, where neither the data-generation process nor the true coefficient values are known. Second, it demonstrates how EnsembleIV can be combined with modern deep-learning techniques for language tasks. Finally, because the underlying data-generation process is unknown, there are no guarantees that the assumptions behind EnsembleIV (as discussed in Section \ref{sec:theory_IV}) will hold. We use this real-world dataset as a testbed to develop a data-driven diagnostic procedure that can assess the satisfaction of Assumption II.\footnote{Note that we still maintain Assumption I, i.e., the mean independence of the regression error term with respect to regression covariates, in the absence of measurement error. If this assumption fails, it implies that other types of endogeneity exist in the regression that go beyond our focus.}

\subsection{Evaluation Setup and Results} \label{section:evaluation_facebook_results}
We use the dataset collected by \cite{yang2019understanding}, which includes a total of 429,015 user-generated posts from 2012, appearing on the business pages of 41 Fortune 500 companies. To build a sentiment classifier, a random subset of 10,157 posts had been manually labeled by 5 independent Amazon Mechanical Turk workers, and their ground truth sentiment labels (positive or negative) were determined via majority voting (i.e., $|D_{label}| = 10157$). The remaining 418,858 unlabeled posts make up $D_{unlabel}$.

For the first-phase machine learning task, we train a binary classifier to predict the sentiment of each post. We consider two different machine learning techniques when building the sentiment classifier. In the first approach, which we refer to as ``BoW + RF", we adopt a traditional bag-of-words representation for posts, namely a term-frequency inverse document-frequency (TF-IDF) matrix.\footnote{We follow standard text pre-processing steps to construct the TF-IDF matrix, including tokenization, punctuation removal, lower-casing, and stop-word removal. We further remove words that are too frequent or too infrequent (showing up in more than 10,000 documents or fewer than 50 documents). The final vocabulary contains 785 unique words.} We train a random forest classifier comprised of 100 trees, taking the TF-IDF matrix as input. In the second approach, which we refer to as ``BERT + RF", we encode each post as a 768-dimension embedding vector using the BERT pre-trained model \citep{devlin2018bert}. As we have learned from the sensitivity analyses in Section \ref{sec:sensitivity_performance}, it is beneficial to fine-tune the first-phase model and also restrict the number of folds in cross-fitting. Therefore, we consider three different ensemble techniques (random forest, XGBoost, and LightGBM) and fine-tune the number of individual learners in each model via 3-fold cross-validation. In this application, the LightGBM technique \citep{ke2017lightgbm} ultimately outperforms random forest and XGBoost. We summarize the accuracy and F-scores of the best-performing models in Table \ref{table:facebook_classifier}.

\begin{table}[!tbh]
    \centering 
    \caption{Predictive Performance of Sentiment Classifiers (Obtained via 3-Fold Cross-Validation)} \label{table:facebook_classifier} 
    \begin{tabular}{c|c|c}
    \hline
         &  BoW + LightGBM ($M = 150$)  &  BERT + LightGBM ($M = 100$) \\
    \hline
    Accuracy  &  0.867  &  0.844  \\
    F-score of ``Positive" Class  &  0.744  &  0.676 \\
    F-score of ``Negative" Class  &  0.910  &  0.897 \\
    \hline
    \end{tabular}
\end{table}

For the second-phase inference task, we estimate a linear regression on $D_{unlabel}$ with the number of comments received by each post serving as the dependent variable, log-transformed to reduce skewness. The sentiment of each post, denoted as $sentiment$, takes a value of 1 if the predicted sentiment is positive and 0 otherwise. This variable serves as the mismeasured explanatory variable. In addition, we control for several other covariates, including (i) the (log-transformed) number of words in a post ($wordcount$) and (ii) the content type of the post ($type$, one of photo, status, video, or link). In Table \ref{table:facebook}, we report estimation results from the Biased regression, Unbiased regression, and EnsembleIV, all with bootstrapped standard errors. For EnsembleIV, we apply 3-fold cross-fitting and the PCA-based IV selection (the other IV selection approaches produce similar results). Note that the unbiased estimates do not rely on machine learning predictions and thus remain fixed across results associated with each of the two sentiment classifiers.

\begin{table}[!tbh]
    \centering 
    \caption{Evaluation Results on Facebook Data} \label{table:facebook} 
    \begin{tabular}{c|c | c c | c c}
    \hline
      &  & \multicolumn{2}{c|}{BoW + RF}  &  \multicolumn{2}{c}{BERT + RF}  \\
    \hline
      &  Unbiased  &  Biased  &  Ens.IV &   Biased  &  Ens.IV  \\
    \hline
    \textbf{sentiment} & \textbf{$-$0.202$^{***}$} & \textbf{$-$0.151$^{***}$} & \textbf{$-$0.217$^{***}$} & \textbf{$-$0.126$^{***}$} & \textbf{$-$0.206$^{***}$} \\ 
      & (0.016) & (0.004) & (0.008) & (0.005) & (0.011) \\ 
     wordcount & 0.142$^{***}$ & 0.189$^{***}$ & 0.181$^{***}$ & 0.187$^{***}$ & 0.181$^{***}$ \\ 
      & (0.007) & (0.001) & (0.002) & (0.001) & (0.002) \\ 
     type\_photo & 0.317$^{***}$ & 0.439$^{***}$ & 0.440$^{***}$ & 0.437$^{***}$ & 0.440$^{***}$ \\ 
      & (0.055) & (0.011) & (0.011) & (0.011) & (0.011) \\ 
     type\_status & 0.429$^{***}$ & 0.282$^{***}$ & 0.278$^{***}$ & 0.284$^{***}$ & 0.282$^{***}$ \\ 
      & (0.036) & (0.008) & (0.008) & (0.008) & (0.009) \\ 
     type\_video & $-$0.023 & 0.048$^{*}$ & 0.048$^{*}$ & 0.042 & 0.045$^{*}$ \\ 
      & (0.126) & (0.022) & (0.022) & (0.022) & (0.022) \\ 
     Constant & $-$0.024 & 0.061$^{***}$ & 0.133$^{***}$ & 0.053$^{***}$ & 0.128$^{***}$ \\ 
      & (0.039) & (0.009) & (0.010) & (0.009) & (0.011) \\
    \hline
    \end{tabular}
\end{table}

Although the true coefficient values are unknown, we can nevertheless use the unbiased estimates as a reasonable baseline. Several findings are worth noting. First, directly using the aggregated predictions from LightGBM in the second-phase regressions results in nontrivial bias. The magnitude of the coefficient on $sentiment$ is significantly underestimated. Second, EnsembleIV can mitigate the estimation bias on the sentiment variable in both cases, producing point estimates on $sentiment$ that are close to the unbiased value and with smaller standard errors. This again demonstrates EnsembleIV's utility to reduce bias compared to the biased estimate and improve precision compared to the unbiased estimate. 

\subsection{Diagnostic Procedure} \label{section:evaluation_facebook_diagnostic}
Recall from Section \ref{sec:theory_IV} that the validity of EnsembleIV relies on two assumptions regarding the error term of the second-phase regression, $\varepsilon$, namely that (i) in the absence of measurement error, $\varepsilon$ is mean independent of regression covariates, and (ii) conditional on the regression covariates, $\varepsilon$ is uncorrelated with the prediction errors of all individual learners. While Assumption I simply means that there are no sources of endogeneity other than measurement error, Assumption II is needed to establish the validity of using (transformed) individual learners as IVs. Recent work has begun to consider scenarios in which Assumption II is violated \citep{wei2022unstructured,allon2023machine}. This can happen when the machine learning models exploit ``peripheral features" of inputs that affect $Y$ but are not included in $X$ or $\boldsymbol{W}$. 

We develop a data-driven diagnostic procedure that a researcher can employ to evaluate whether this peripheral feature challenge is likely to be present in the variables generated by an ensemble learning technique on a particular dataset. The procedure uses a subset of the labeled data to evaluate the correlations between individual learners' prediction errors and the residuals of the measurement-error-free regression (as an unbiased and consistent estimate of $\varepsilon$). Specifically, the diagnostic is designed as follows:
\begin{enumerate}
    \item Given a labeled dataset $D_{label}$, we reserve a random partition $D_{diagnostic}$ and use the remaining data in $D_{label} \setminus D_{diagnostic}$ for training the first-phase ensemble ML model and estimating the $\lambda$ parameters for IV transformation (via standard sample-splitting);
    \item Estimate the Unbiased OLS regression on $D_{label} \setminus D_{diagnostic}$;
    \item Apply the ensemble ML model on $D_{diagnostic}$ to obtain predictions and apply Algorithm 1 to obtain transformed IVs (essentially treating $D_{diagnostic}$ as $D_{unlabel}$ in Algorithm 1);
    \item Apply the estimated OLS regression on $D_{diagnostic}$ to obtain residuals $r_{diagnostic}$;
    \item For any given individual learner $\widehat{X}^{(i)}$ designated as the endogenous covariate and its corresponding transformed IVs $\widetilde{Z}^{(j)}$, calculate the empirical correlation $Corr(\widetilde{Z}^{(j)} - X, r_{diagnostic})$ and test for a statistically significant association.
\end{enumerate}
Because $D_{diagnostic}$ is a random sample that shares the same distribution as the actual $D_{unlabel}$, the estimated $Corr(\widetilde{Z}^{(j)} - X, r_{diagnostic})$ contains testable evidence whether  generated and transformed IVs in this dataset tend to exhibit correlation with the regression error term. In practice, researchers can select their preferred test of correlation and significance level to assess whether the empirical correlation is statistically significant. Here we provide one exemplary illustration of how this could be done. Let $\widetilde{Z}^{(j)}_{i}$ denote a transformed IV for endogenous covariate $\widehat{X}^{(i)}$ ($i \neq j$). Given an ensemble model of $M$ individual learners, the EnsembleIV procedure uses $M(M-1)$ such transformed IVs (i.e., $M-1$ transformed IVs for each of the $M$ individual learners designated as the endogenous covariate). One could construct a joint test of all $M(M-1)$ correlations, which offers an empirical assessment of how severe the challenge of ``peripheral features" may be. This joint test is illustrated in the following hypothesis:
\begin{equation}
\label{eq:hypo}
\begin{split}
    H_0: & \forall i \in \{1, \ldots, M\} \text{ and } i \neq j, \mathbb{E}(Corr(\widetilde{Z}^{(j)}_{i} - X, r_{diagnostic})) = 0 \\
    H_1: & \exists i \in \{1, \ldots, M\} \text{ and } i \neq j, \mathbb{E}(Corr(\widetilde{Z}^{(j)}_{i} - X, r_{diagnostic})) \neq 0
\end{split}
\end{equation}
One example test statistic that can be used to test the above hypothesis is:
\begin{equation}
\label{eq:test_stat}
TS = \frac{1}{M(M-1)} \sum_{i \neq j} \left|Corr(\widetilde{Z}^{(j)}_{i} - X, r_{diagnostic}) \right|
\end{equation}
i.e., the average absolute correlation between the error of a transformed IV and the residual term. To perform a hypothesis test, we would need the distribution of this test statistic under the null of zero correlation. Deriving the theoretical sampling distribution of this test statistic is challenging. However, we can rely on a standard \textit{permutation test} \citep{welch1990construction} to approximate the distribution. Specifically, we permute the ordering of records in $r_{diagnostic}$ and re-compute the test statistic. Over a large number of permutations, this procedure generates an empirical distribution of the test statistic under the null of zero correlation. Then, we can compute the empirical $p$-value by comparing the observed test statistic from the original (``un-permuted") sample to this empirical distribution.\footnote{Note that the procedure we propose considers forming a single sample of $D_{diagnostic}$, providing a single $TS$ to test. However, Type II errors can be decreased with a $K$-fold sample-splitting procedure, similar to what we implemented in the ``cross-fitting'' enhanced version of Ensemble IV. Specifically, $D_{label}$ can be randomly partitioned into $K$ folds and, in each iteration, a unique fold is utilized as $D_{diagnostic}$, producing $K$ independent test statistics and associated $p$-values. The $p$-values from these tests can be combined using standard methods \citep[e.g., Fisher's Method,][]{fishers-method} to leverage the full $D_{label}$ for testing.}

Should the null hypothesis of zero correlation be rejected (i.e., the empirical $p$-value falls below the significance threshold chosen by the researcher), indicating a nontrivial peripheral feature challenge, the researcher will need to take steps to address the issue. For this purpose, we develop an extension to EnsembleIV that explicitly accounts for the potential correlation between prediction errors and peripheral features (see Appendix \ref{ap:extension} for details). That said, if the null hypothesis cannot be rejected, then the researcher may have some confidence that the peripheral features challenge is absent or is too weak to be detectable in the given empirical context (and may, therefore, not pose a significant threat to estimation and inference). In this case, the researcher can proceed by applying the standard form of EnsembleIV, as presented earlier.

We carry out the diagnostic procedure on the Facebook dataset. Among 10,157 labeled data, we reserve 1,000 as $D_{diagnostic}$. We train a LightGBM model with the BERT representation because model evaluation shows that it produces \textit{worse} performance than the bag-of-words representation (perhaps the model has missed some important features, making the ``peripheral features" challenge more likely to be present in this setting). We provide a descriptive summary of the empirical correlations in Table \ref{table:fb_diagnostic} below.

\begin{table}[!tbh]
    \centering 
    \caption{Descriptive Statistics of Correlations Before and After IV Transformation (Facebook Data)} \label{table:fb_diagnostic}
    \begin{tabular}{c | c | c}
    \hline
    &   Before IV transformation  & After IV transformation  \\
    &   $|Corr(\widehat{X}^{(j)} - X, r_{diagnostic})|$  &  $|Corr(\widetilde{Z}^{(j)} - X, r_{diagnostic})|$ \\
    \hline
    Min  &  0.00003  &  0.00006 \\
    Max  &  0.032  &  0.058 \\
    Mean &  0.022  &  0.016$^{***}$ \\
    SD   &  0.009  &  0.008  \\
    \hline
    \end{tabular}
    \caption*{\textit{Note}. $^{***} p < 0.001$ based on a paired $t$-test.}
\end{table}

We find that the correlations between individual learners' prediction errors and the regression residual are small even before IV transformation, potentially indicating that the BERT representation already captures a sufficient amount of textual information. After IV transformation, the correlations become even weaker -- a paired $t$-test shows significantly weaker correlations after IV transformation ($p < 0.001$). Applying the permutation test discussed above, we fail to reject the null hypothesis of zero correlation ($TS = 0.016, p = 0.5982$, computed based on 10,000 permutations). While this is not confirmation that the null is true, it does indicate that the empirical correlations we observed are consistent with those we might expect under zero correlation. Therefore, the problem of ``peripheral features" appears sufficiently negligible in this specific dataset and using the original EnsembleIV method is reasonable.

In addition to the above experiment on the Facebook data, we have constructed a set of simulations that help demonstrate that the proposed diagnostic procedure is useful for detecting invalid IVs when there are known strong correlations with peripheral features. Details of these simulations are provided in Appendix \ref{ap:diagnostic}. 

Finally, we acknowledge that more research is needed to fully understand (1) the prevalence and challenge of ``peripheral features" in practice and (2) how robust different correction methods (EnsembleIV included) are to this challenge. Recent work including \cite{wei2022unstructured} and \cite{allon2023machine} represent laudable steps towards a better understanding of these issues, and we expect the literature will continue proposing approaches that expand the ability to perform inference in the presence of different degrees of the ``peripheral features" challenge.

\section{Discussion}
Despite increasing popularity in empirical studies, the integration of machine-learning-generated variables into regression models for statistical inference suffers from a measurement error problem, which can bias estimation and threaten the validity of inference. In this paper, we develop a novel approach to alleviate associated estimation biases. Our proposed approach, EnsembleIV, creates valid and strong instrumental variables from individual learners in an ensemble model and uses them to obtain corrected estimates that are robust to measurement error. 

The core idea behind EnsembleIV is that individual learners in an ensemble (e.g., individual trees from a random forest) typically produce correlated predictions. As a result, for a specific individual learner, predictions from other individual learners can serve as relevant but not perfectly excluded instrumental variables. Furthermore, the technique based on the work of Nevo and Rosen \citep{nevo2012identification} enables the transformation of these candidate IVs into (asymptotically) valid ones. Our proposed EnsembleIV procedure follows this idea to generate, transform, and select IVs in a systematic and data-driven manner and use them to correct estimation bias arising from the measurement error.

Our empirical evaluations, using both synthetic and real-world datasets, show that EnsembleIV can effectively reduce estimation biases for both linear and generalized linear regression specifications, and its effectiveness is not restricted by the type of machine-learning-generated variable (binary or continuous) or the specific ensemble learning technique (random forest or boosting). For unstructured data (e.g., text), EnsembleIV can be easily used in conjunction with representations learned by deep learning networks. Compared with several alternative bias correction methods, including ForestIV \citep{yang2022achieving}, regression calibration \citep{gleser1992importance}, and GMM \citep{fong2021machine}, EnsembleIV can improve the estimation efficiency over regression calibration and ForestIV, while offering greater flexibility than GMM (i.e., EnsembleIV does not need regression-specific adjustments). Moreover, decreased predictive performance of the first-phase ensemble learning model (e.g., due to having a harder predictive task or fewer predictive features) translates to wider confidence intervals, but not increased biases, of EnsembleIV estimates. Overall, EnsembleIV represents a versatile and highly practical tool for drawing robust statistical inference with machine-learning-generated variables. 

Our work also opens up several directions for future research. As shown in our sensitivity analyses with ForestIV, it appears that EnsembleIV's correction effectiveness is best when the size of the labeled data partition is at least moderately large. Future work may seek to improve the small sample performance of EnsembleIV -- this can enhance EnsembleIV's usefulness in settings where it is simply infeasible to obtain more than a small amount of labeled data. Further, recognizing the emerging popularity of Double/Debiased Machine Learning \citep[DML,][]{chernozhukov2018double}, it would also be interesting to combine EnsembleIV with DML from both theoretical and empirical perspectives. 

\bibliography{manuscript.bbl}

\begin{thebibliography}{}

\bibitem[Allon et~al., 2023]{allon2023machine}
Allon, G., Chen, D., Jiang, Z., and Zhang, D. (2023).
\newblock Machine learning and prediction errors in causal inference.
\newblock {\em Available at SSRN 4480696}.

\bibitem[Andrews et~al., 2019]{andrews2019weak}
Andrews, I., Stock, J.~H., and Sun, L. (2019).
\newblock Weak instruments in instrumental variables regression: Theory and practice.
\newblock {\em Annual Review of Economics}, 11:727--753.

\bibitem[Angrist and Pischke, 2008]{angrist2008mostly}
Angrist, J.~D. and Pischke, J.-S. (2008).
\newblock {\em Mostly harmless econometrics: An empiricist's companion}.
\newblock Princeton university press.

\bibitem[Belloni et~al., 2012]{belloni2012sparse}
Belloni, A., Chen, D., Chernozhukov, V., and Hansen, C. (2012).
\newblock Sparse models and methods for optimal instruments with an application to eminent domain.
\newblock {\em Econometrica}, 80(6):2369--2429.

\bibitem[Belloni et~al., 2017]{belloni2017high-inference}
Belloni, A., Chernozhukov, V., Fernández-Val, I., and Hansen, C. (2017).
\newblock Program evaluation and causal inference with high-dimensional data.
\newblock {\em Econometrica}, 85(1):233--298.

\bibitem[Belloni et~al., 2014]{belloni2014post-inference}
Belloni, A., Chernozhukov, V., and Hansen, C. (2014).
\newblock Inference on treatment effects after selection among high-dimensional controls.
\newblock {\em The Review of Economic Studies}, 81(2):608--650.

\bibitem[Belloni et~al., 2016]{belloni2016post-inference-glm}
Belloni, A., Chernozhukov, V., and Wei, Y. (2016).
\newblock Post-selection inference for generalized linear models with many controls.
\newblock {\em Journal of Business \& Economic Statistics}, 34(4):606--619.

\bibitem[Berk et~al., 2013]{berk2013valid}
Berk, R., Brown, L., Buja, A., Zhang, K., and Zhao, L. (2013).
\newblock Valid post-selection inference.
\newblock {\em The Annals of Statistics}, 41(2):802--837.

\bibitem[Bound et~al., 1995]{bound1995problems}
Bound, J., Jaeger, D.~A., and Baker, R.~M. (1995).
\newblock Problems with instrumental variables estimation when the correlation between the instruments and the endogenous explanatory variable is weak.
\newblock {\em Journal of the American statistical association}, 90(430):443--450.

\bibitem[Breiman, 2001]{Breiman2001}
Breiman, L. (2001).
\newblock {Random forests}.
\newblock {\em Machine Learning}, 45(1):5--32.

\bibitem[Buzas and Stefanski, 1996]{buzas1996instrumental}
Buzas, J.~S. and Stefanski, L.~A. (1996).
\newblock Instrumental variable estimation in generalized linear measurement error models.
\newblock {\em Journal of the American Statistical Association}, 91(435):999--1006.

\bibitem[Carroll et~al., 1995]{carroll1995measurement}
Carroll, R.~J., Ruppert, D., and Stefanski, L.~A. (1995).
\newblock {\em Measurement error in nonlinear models}, volume 105.
\newblock CRC press.

\bibitem[Carroll et~al., 2006]{carroll2006measurement}
Carroll, R.~J., Ruppert, D., Stefanski, L.~A., and Crainiceanu, C.~M. (2006).
\newblock {\em Measurement error in nonlinear models: a modern perspective}.
\newblock Chapman and Hall/CRC.

\bibitem[Cengiz et~al., 2022]{cengiz2022seeing}
Cengiz, D., Dube, A., Lindner, A., and Zentler-Munro, D. (2022).
\newblock Seeing beyond the trees: Using machine learning to estimate the impact of minimum wages on labor market outcomes.
\newblock {\em Journal of Labor Economics}, 40(S1):S203--S247.

\bibitem[Chen and Guestrin, 2016]{chen2016xgboost}
Chen, T. and Guestrin, C. (2016).
\newblock Xgboost: A scalable tree boosting system.
\newblock In {\em Proceedings of the 22nd acm sigkdd international conference on knowledge discovery and data mining}, pages 785--794.

\bibitem[Chernozhukov et~al., 2018]{chernozhukov2018double}
Chernozhukov, V., Chetverikov, D., Demirer, M., Duflo, E., Hansen, C., Newey, W., and Robins, J. (2018).
\newblock Double/debiased machine learning for treatment and structural parameters: Double/debiased machine learning.
\newblock {\em The Econometrics Journal}, 21(1).

\bibitem[Chernozhukov et~al., 2015]{chernozhukov2015post-inference}
Chernozhukov, V., Hansen, C., and Spindler, M. (2015).
\newblock Post-selection and post-regularization inference in linear models with many controls and instruments.
\newblock {\em American Economic Review}, 105(5):486--90.

\bibitem[Conley et~al., 2012]{conley2012plausibly}
Conley, T.~G., Hansen, C.~B., and Rossi, P.~E. (2012).
\newblock Plausibly exogenous.
\newblock {\em Review of Economics and Statistics}, 94(1):260--272.

\bibitem[Davidson and MacKinnon, 2006]{davidson2006boostrap}
Davidson, R. and MacKinnon, J.~G. (2006).
\newblock The power of bootstrap and asymptotic tests.
\newblock {\em Journal of Econometrics}, 133(2):421--441.

\bibitem[Davidson and MacKinnon, 2008]{davidson2008bootstrap}
Davidson, R. and MacKinnon, J.~G. (2008).
\newblock Bootstrap inference in a linear equation estimated by instrumental variables.
\newblock {\em The Econometrics Journal}, 11(3):443--477.

\bibitem[Devlin et~al., 2018]{devlin2018bert}
Devlin, J., Chang, M.-W., Lee, K., and Toutanova, K. (2018).
\newblock Bert: Pre-training of deep bidirectional transformers for language understanding.
\newblock {\em arXiv preprint arXiv:1810.04805}.

\bibitem[Donald and Newey, 2001]{donald2001choosing}
Donald, S.~G. and Newey, W.~K. (2001).
\newblock Choosing the number of instruments.
\newblock {\em Econometrica}, 69(5):1161--1191.

\bibitem[Fanaee-T and Gama, 2014]{fanaee2014event}
Fanaee-T, H. and Gama, J. (2014).
\newblock Event labeling combining ensemble detectors and background knowledge.
\newblock {\em Progress in Artificial Intelligence}, 2(2-3):113--127.

\bibitem[Fisher, 1925]{fishers-method}
Fisher, R.~A. (1925).
\newblock {\em Statistical methods for research workers}.
\newblock Oliver and Boyd, 5th edition.

\bibitem[Fong and Tyler, 2021]{fong2021machine}
Fong, C. and Tyler, M. (2021).
\newblock Machine learning predictions as regression covariates.
\newblock {\em Political Analysis}, 29(4):467--484.

\bibitem[Gleser, 1992]{gleser1992importance}
Gleser, L.~J. (1992).
\newblock The importance of assessing measurement reliability in multivariate regression.
\newblock {\em Journal of the American Statistical Association}, 87(419):696--707.

\bibitem[Goh et~al., 2013]{goh2013social}
Goh, K.-Y., Heng, C.-S., and Lin, Z. (2013).
\newblock Social media brand community and consumer behavior: Quantifying the relative impact of user-and marketer-generated content.
\newblock {\em Information systems research}, 24(1):88--107.

\bibitem[Greene, 2003]{greene2003econometric}
Greene, W.~H. (2003).
\newblock {\em Econometric analysis}.
\newblock Pearson Education India.

\bibitem[Hall, 1992]{hall1992bootstrap}
Hall, P. (1992).
\newblock {\em The Bootstrap and Edgeworth Expansion}.
\newblock Springer, New York.

\bibitem[Hopkins and King, 2007]{hopkins2007extracting}
Hopkins, D. and King, G. (2007).
\newblock Extracting systematic social science meaning from text.
\newblock {\em Manuscript available at http://gking. harvard. edu/files/words. pdf}, 20(07).

\bibitem[Horowitz, 2019]{horowitz2019bootstrap}
Horowitz, J.~L. (2019).
\newblock Bootstrap methods in econometrics.
\newblock {\em Annual Review of Economics}, 11(1):193--224.

\bibitem[Hu and Schennach, 2008]{hu2008instrumental}
Hu, Y. and Schennach, S.~M. (2008).
\newblock Instrumental variable treatment of nonclassical measurement error models.
\newblock {\em Econometrica}, 76(1):195--216.

\bibitem[Ke et~al., 2017]{ke2017lightgbm}
Ke, G., Meng, Q., Finley, T., Wang, T., Chen, W., Ma, W., Ye, Q., and Liu, T.-Y. (2017).
\newblock Lightgbm: A highly efficient gradient boosting decision tree.
\newblock {\em Advances in neural information processing systems}, 30.

\bibitem[K{\"{u}}chenhoff et~al., 2006]{Kuchenhoff2006}
K{\"{u}}chenhoff, H., Mwalili, S.~M., and Lesaffre, E. (2006).
\newblock {A general method for dealing with misclassification in regression: The misclassification SIMEX}.
\newblock {\em Biometrics}, 62(1):85--96.

\bibitem[Lee et~al., 2018]{lee2018advertising}
Lee, D., Hosanagar, K., and Nair, H.~S. (2018).
\newblock Advertising content and consumer engagement on social media: Evidence from facebook.
\newblock {\em Management Science}, 64(11):5105--5131.

\bibitem[Mehrhoff, 2009]{mehrhoff2009solution}
Mehrhoff, J. (2009).
\newblock A solution to the problem of too many instruments in dynamic panel data gmm.
\newblock {\em Available at SSRN 2785360}.

\bibitem[Moreno and Terwiesch, 2014]{moreno2014doing}
Moreno, A. and Terwiesch, C. (2014).
\newblock Doing business with strangers: Reputation in online service marketplaces.
\newblock {\em Information Systems Research}, 25(4):865--886.

\bibitem[Moro et~al., 2014]{moro2014data}
Moro, S., Cortez, P., and Rita, P. (2014).
\newblock A data-driven approach to predict the success of bank telemarketing.
\newblock {\em Decision Support Systems}, 62:22--31.

\bibitem[Murray, 2006]{murray2006avoiding}
Murray, M.~P. (2006).
\newblock Avoiding invalid instruments and coping with weak instruments.
\newblock {\em Journal of economic Perspectives}, 20(4):111--132.

\bibitem[Nevo and Rosen, 2012]{nevo2012identification}
Nevo, A. and Rosen, A.~M. (2012).
\newblock Identification with imperfect instruments.
\newblock {\em Review of Economics and Statistics}, 94(3):659--671.

\bibitem[Oxley and McAleer, 1993]{oxley1993econometric}
Oxley, L. and McAleer, M. (1993).
\newblock Econometric issues in macroeconomic models with generated regressors.
\newblock {\em Journal of Economic Surveys}, 7(1):1--40.

\bibitem[Pagan, 1984]{pagan1984econometric}
Pagan, A. (1984).
\newblock Econometric issues in the analysis of regressions with generated regressors.
\newblock {\em International Economic Review}, 25(1):221--247.

\bibitem[Qiao and Huang, 2021]{qiao2021correcting}
Qiao, M. and Huang, K.-W. (2021).
\newblock Correcting misclassification bias in regression models with variables generated via data mining.
\newblock {\em Information Systems Research}, 32(2):462--480.

\bibitem[Roodman, 2009]{roodman2009note}
Roodman, D. (2009).
\newblock A note on the theme of too many instruments.
\newblock {\em Oxford Bulletin of Economics and Statistics}, 71(1):135--158.

\bibitem[Stefanski and Cook, 1995]{Stefanski1995}
Stefanski, A. L.~A. and Cook, J.~R. (1995).
\newblock {Simulation-Extrapolation : The Measurement Error Jackknife}.
\newblock {\em Journal of the American Statistical Association}, 90(432):1247--1256.

\bibitem[Stock et~al., 2002]{stock2002survey}
Stock, J.~H., Wright, J.~H., and Yogo, M. (2002).
\newblock A survey of weak instruments and weak identification in generalized method of moments.
\newblock {\em Journal of Business \& Economic Statistics}, 20(4):518--529.

\bibitem[Stock and Yogo, 2002]{stock2002testing}
Stock, J.~H. and Yogo, M. (2002).
\newblock Testing for weak instruments in linear iv regression.
\newblock {\em National Bureau of Economic Research Cambridge}.

\bibitem[Terza et~al., 2008]{terza2008two}
Terza, J.~V., Basu, A., and Rathouz, P.~J. (2008).
\newblock Two-stage residual inclusion estimation: addressing endogeneity in health econometric modeling.
\newblock {\em Journal of health economics}, 27(3):531--543.

\bibitem[Tirunillai and Tellis, 2012]{tirunillai2012does}
Tirunillai, S. and Tellis, G.~J. (2012).
\newblock Does chatter really matter? dynamics of user-generated content and stock performance.
\newblock {\em Marketing Science}, 31(2):198--215.

\bibitem[Wan et~al., 2018]{wan2018general}
Wan, F., Small, D., and Mitra, N. (2018).
\newblock A general approach to evaluating the bias of 2-stage instrumental variable estimators.
\newblock {\em Statistics in medicine}, 37(12):1997--2015.

\bibitem[Wang et~al., 2020]{wang2020methods}
Wang, S., McCormick, T.~H., and Leek, J.~T. (2020).
\newblock Methods for correcting inference based on outcomes predicted by machine learning.
\newblock {\em Proceedings of the National Academy of Sciences}, 117(48):30266--30275.

\bibitem[Wei and Malik, 2022]{wei2022unstructured}
Wei, Y. and Malik, N. (2022).
\newblock Unstructured data, econometric models, and estimation bias.
\newblock {\em SSRN}.

\bibitem[Welch, 1990]{welch1990construction}
Welch, W.~J. (1990).
\newblock Construction of permutation tests.
\newblock {\em Journal of the American Statistical Association}, 85(411):693--698.

\bibitem[Wooldridge, 2002]{wooldridge2002}
Wooldridge, J.~M. (2002).
\newblock {\em Econometric analysis of cross section and panel data}.
\newblock MIT Press, Cambridge and London.

\bibitem[Wu et~al., 2020]{wu2020air}
Wu, X., Nethery, R.~C., Sabath, M.~B., Braun, D., and Dominici, F. (2020).
\newblock Air pollution and covid-19 mortality in the united states: Strengths and limitations of an ecological regression analysis.
\newblock {\em Science advances}, 6(45):eabd4049.

\bibitem[Yang et~al., 2018]{yang2018mind}
Yang, M., Adomavicius, G., Burtch, G., and Ren, Y. (2018).
\newblock Mind the gap: Accounting for measurement error and misclassification in variables generated via data mining.
\newblock {\em Information Systems Research}, 29(1):4--24.

\bibitem[Yang et~al., 2022]{yang2022achieving}
Yang, M., McFowland~III, E., Burtch, G., and Adomavicius, G. (2022).
\newblock Achieving reliable causal inference with data-mined variables: A random forest approach to the measurement error problem.
\newblock {\em INFORMS Journal on Data Science}, 1(2).

\bibitem[Yang et~al., 2019]{yang2019understanding}
Yang, M., Ren, Y., and Adomavicius, G. (2019).
\newblock Understanding user-generated content and customer engagement on facebook business pages.
\newblock {\em Information Systems Research}, 30(3):839--855.

\bibitem[Zhang et~al., 2023]{zhang2023debiasing}
Zhang, J., Xue, W., Yu, Y., and Tan, Y. (2023).
\newblock Debiasing machine-learning-or ai-generated regressors in partial linear models.
\newblock {\em Available at SSRN}.

\bibitem[Zhang et~al., 2021]{zhang2021makes}
Zhang, S., Lee, D., Singh, P.~V., and Srinivasan, K. (2021).
\newblock What makes a good image? airbnb demand analytics leveraging interpretable image features.
\newblock {\em Management Science}, 68(8).

\end{thebibliography}

\clearpage
\appendix

\section{Descriptive Evidence of IV Transformation and IV selection Effectiveness} \label{ap:descriptive}
\begin{figure}[tbph]
    \centering
    \includegraphics[width=\linewidth]{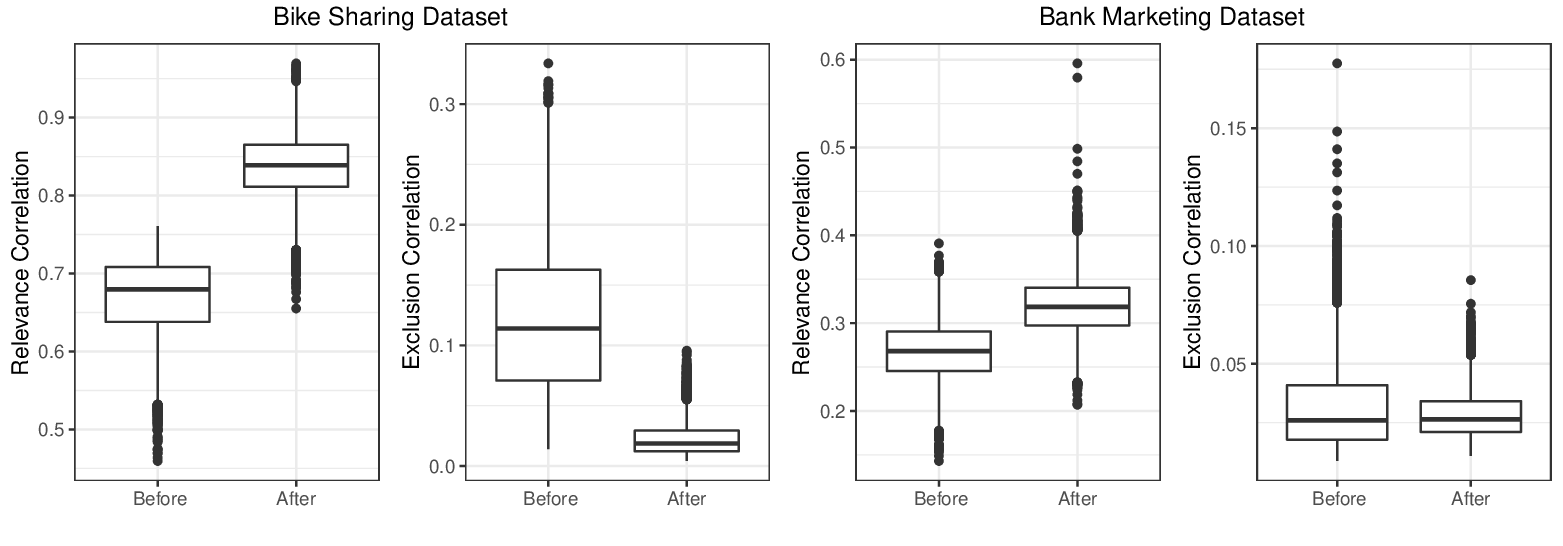}
    \caption{Average Relevance/Exclusion Measures Before and After IV Transformation and Selection}
    \label{fig:before_after}
\end{figure}

Using both the Bike Sharing and Bank Marketing datasets, we first provide descriptive evidence to demonstrate that the IV transformation step and IV selection step (using LASSO selection for illustration) can indeed create (approximately) valid instruments and select strong ones. In particular, for a given mismeasured covariate $\widehat{X}^{(i)}$, denote $\Psi_{before}^{(i)} = \{\widehat{X}^{(j)}\}_{j \neq i}$ as the set of candidate IVs before any transformation or selection, and denote $\Psi_{after}^{(i)} \subseteq \{\widetilde{Z}^{(j)}\}_{j \neq i}$ as the set of transformed IVs selected by the LASSO step. We then compute $\frac{1}{|\Psi_{\bullet}^{(i)}|} \sum_{z \in \Psi_{\bullet}^{(i)}} |Corr(\widehat{X}^{(i)}, z)|$ to measure the average relevance of selected IVs, and $\frac{1}{|\Psi_{\bullet}^{(i)}|} \sum_{z \in \Psi_{\bullet}^{(i)}} |Corr(\widehat{X}^{(i)} - X, z)|$ to measure the average exclusion of selected IVs. In Figure \ref{fig:before_after}, we plot the distribution of the average relevance and exclusion measures $\forall i \in \{1, \ldots, 100\}$ and across all simulation runs, both before and after the transformation and selection steps.

On the Bike Sharing dataset, the average relevance and exclusion measures clearly increase and decrease, respectively, after the IV transformation and selection steps. On the Bank Marketing dataset, the average relevance increase is also clear, whereas the average exclusion measure is already small prior to the IV transformation, though its distribution becomes even narrower after the transformation. Overall, these results demonstrate the good performance of IV transformation and selection steps. That said, one can see that the average value of the exclusion measure is not precisely 0 on either dataset, which indicates that even after transformation, the IVs are almost but not perfectly valid. This further emphasizes the practical importance of selecting strong IVs for estimation.\footnote{As Murray \cite[][pg. 128]{murray2006avoiding} notes: ``Strong and almost valid instruments do tend to bias two-stage least squares only a little. However, weak instruments that are almost valid bias two-stage least squares markedly more than do their strong counterparts.''}

\section{Additional Simulations of Diagnostic Procedure} \label{ap:diagnostic}
We simulate the following data generation process:
\begin{equation}
\label{eq:dgp}
    Y = 1 + X + 0.5 \times W + \varepsilon
\end{equation}
where $X \sim N(0,1)$ represents the ground truth values, $W \sim Uniform[0,1]$ represents an exogenous control variable. Importantly, the idiosyncratic error term, $\varepsilon$, is simulated as the sum of a ``peripheral feature" and some exogenous error:
\begin{equation}
\label{eq:error}
    \varepsilon = \underbrace{e_1 + e_2 + \mu}_{\text{Peripheral Feature}} + \underbrace{\tau}_{\text{Exogenous Error}}
\end{equation}
where $e_1, e_2 \sim N(0, \sigma^2)$, $\mu \sim N(0, 0.2^2)$, and $\tau \sim N(0,1)$. We next simulate two mismeasured versions of $X$ (serving as ML-generated variables):
\begin{equation}
\label{eq:pred}
\begin{split}
    X_1 &= X + e_1 + e \\
    X_2 &= X + e_2 + e
\end{split}
\end{equation}
where $e \sim N(0, 0.1^2)$ is a common error component in both variables, which implies that one is only an imperfect instrument for the other (therefore warranting IV transformation). Further, because error components $e_1$ and $e_2$ are present in both $X_1, X_2$ and the peripheral variable, we have introduced correlations between the prediction errors of $X_1, X_2$ and the peripheral variable. By changing the value of $\sigma$, we can alter the strength of the correlations. In the following simulations, we will treat $X_1$ as the endogenous variable and use $X_2$ as the (imperfect) candidate instrument to carry out IV transformation and IV regression. We follow the same permutation test discussed above (with 10,000 permutation runs) to test for correlation between the transformed IV and the residual term.

\begin{figure}[!tbh]
    \centering
    \includegraphics[width=\linewidth]{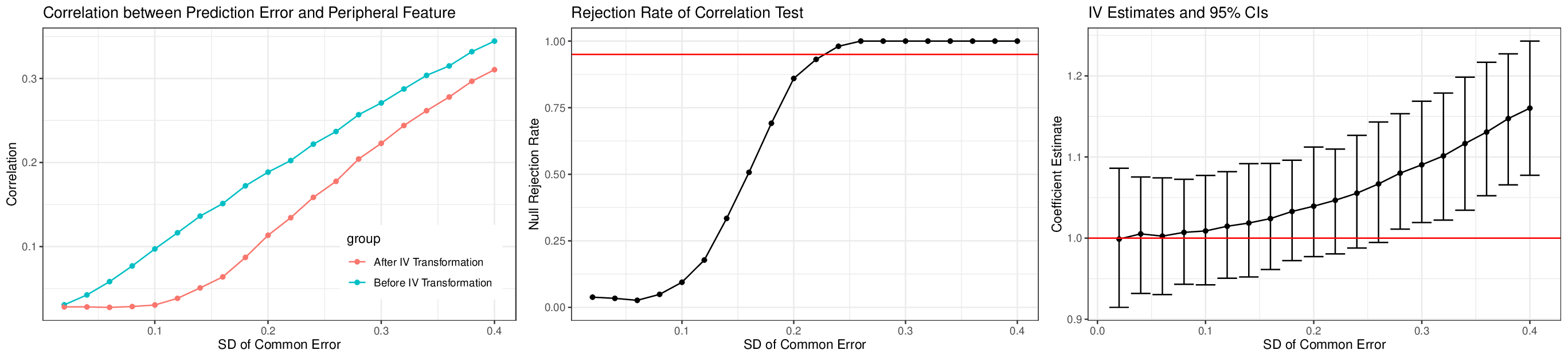}
    \caption{Simulation Results with Varying Strength of Correlation (red line in the second plot indicates 0.95 level; red line in the third plot indicates true coefficient value; results obtained over 500 repetitions)}
    \label{fig:diagnostic_sigma}
\end{figure}

In the first set of simulations, we generate a sample of 5,000 data points and randomly partition it into 3,000, 1,000, and 1,000 observations, for training, testing, and diagnostic, respectively. We vary $\sigma$ across 20 values along the grid $\{0.02, 0.04, \ldots, 0.4\}$ and, under each choice of $\sigma$, repeat the simulations 500 times. We report (i) the correlation between the instrument's prediction error and peripheral variable (calculated based on our diagnostic procedure) both before and after IV transformation (averaged over 500 repetitions); (ii) statistical power -- the rate of rejecting the null of zero correlation (i.e., among 500 repetitions, the proportion of simulation runs that return a rejection result); and (iii) the mean and empirical 95\% confidence interval of the IV estimate (again, obtained over 500 repetitions). The results are presented in Figure \ref{fig:diagnostic_sigma}.

\begin{figure}[!tbh]
    \centering
    \includegraphics[width=\linewidth]{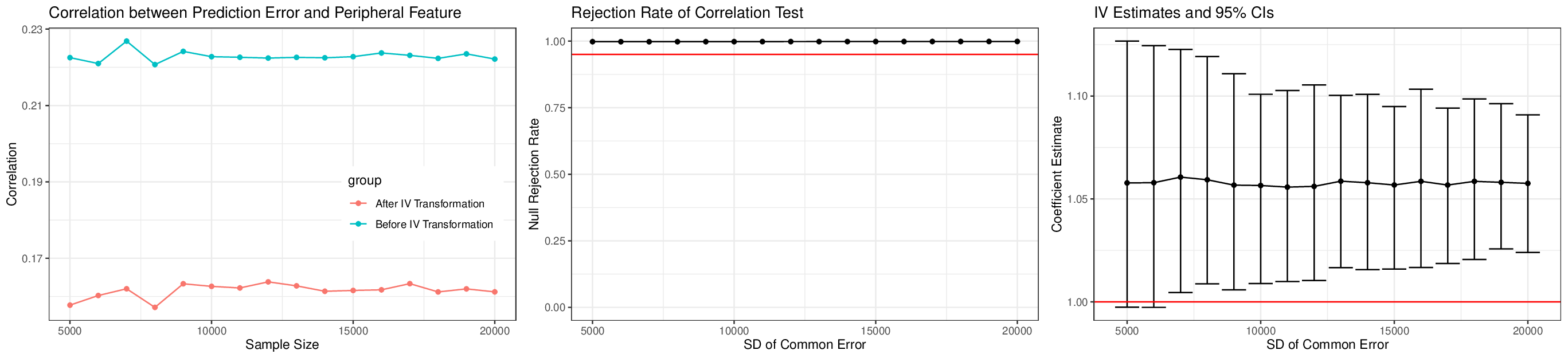}
    \caption{Simulation Results of Varying Sample Size ($\sigma = 0.24$; red line in the second plot indicates 0.95 level; red line in the third plot indicates true coefficient value; results obtained over 500 repetitions)}
    \label{fig:diagnostic_sample}
\end{figure}

Across all values of $\sigma$, our IV transformation step reduces the correlation between prediction error and peripheral feature (statistically significant with $p < 0.001$). When the correlation before transformation is relatively weak (e.g., $\sigma \leq 0.1$, we see that the post-transformation correlation is very close to 0 (which implies a valid IV). In contrast, when the correlation is sufficiently strong (e.g., $\sigma \geq 0.3$) the subsequent bias due to the peripheral feature challenge noticeably impacts IV estimates. Fortunately, we see that our diagnostic test is able to properly detect the issue by rejecting the null of zero correlation. We also observe that when there is a moderate level of correlation (e.g., $\sigma \in (0.24,0.28)$), our diagnostic test also correctly rejects the null of zero correlation even though the IV estimates' confidence intervals still cover the true value in the presence of such non-zero correlation. We suspect these large confidence intervals are a byproduct of the considerable variation in the data-generating process and subsequent uncertainty in the coefficient estimates at smaller sample sizes. Therefore, we conduct a second set of simulations where we fix $\sigma = 0.24$ and vary the total sample size across 16 values along the grid $\{5000, 6000 \ldots, 20000\}$. Under each sample size, we keep the relative proportion of training / testing / diagnostic data to $3:1:1$ and repeat the above analyses. The results, presented in Figure \ref{fig:diagnostic_sample}, show that with larger sample sizes, the estimation bias in IV estimates becomes more evident and the 95\% confidence intervals no longer cover the true value. These two sets of simulations confirm that our diagnostic procedure is reliable, possessing sufficient power against the null of zero correlation, even at moderate levels of correlation.

\section{Extension of EnsembleIV to Account for Peripheral Feature Correlation} \label{ap:extension}
If the diagnostic procedure indicates a significant correlation between prediction errors and peripheral features, then the original EnsembleIV approach would likely yield inconsistent estimates. To deal with this scenario, we next consider an extension to EnsembleIV that explicitly accounts for the potential correlation between prediction errors and peripheral features. Recall the calculation of the $\lambda$ parameter used in IV transformation:
\begin{equation}
\label{eq:lambda_full}
    \lambda = \frac{\rho_{Zu}}{\rho_{\widehat{X}u}} = \frac{Cov(Z,u)}{Cov(\widehat{X},u)} \cdot \frac{\sigma_{\widehat{X}}}{\sigma_Z} = \frac{Cov(Z,\varepsilon) - \beta Cov(Z,e)}{Cov(\widehat{X}, \varepsilon) - \beta Cov(\widehat{X},e)} \cdot \frac{\sigma_{\widehat{X}}}{\sigma_Z}
\end{equation}
In the presence of peripheral feature correlation, $Cov(Z,\varepsilon)$ and $Cov(\widehat{X}, \varepsilon)$ are not zero, but they can still be estimated using $D_{test}$. Similar to the diagnostic procedure, we estimate the unbiased first-phase regression on $D_{train}$, obtain the residuals on $D_{test}$ as a proxy for $\varepsilon$, then compute the empirical correlations between said residuals and individual learner predictions on $D_{test}$. The unknown $\beta$ coefficient can be approximated by the sample estimate on the full labeled data, jointly on $D_{train}$ and $D_{test}$ (i.e., based on the unbiased regression). Together, we calculate a \textit{modified} $\lambda$ value that accounts for the peripheral feature correlations, which can then be used to transform the corresponding IV on $D_{unlabel}$. Please note that cross-fitting can also be applied here, in the same way that it is applied to the original EnsembleIV approach, enabling the full labeled data to eventually be used in estimating $\lambda$.

\begin{figure}[!tbh]
    \centering
    \includegraphics[width=0.5\linewidth]{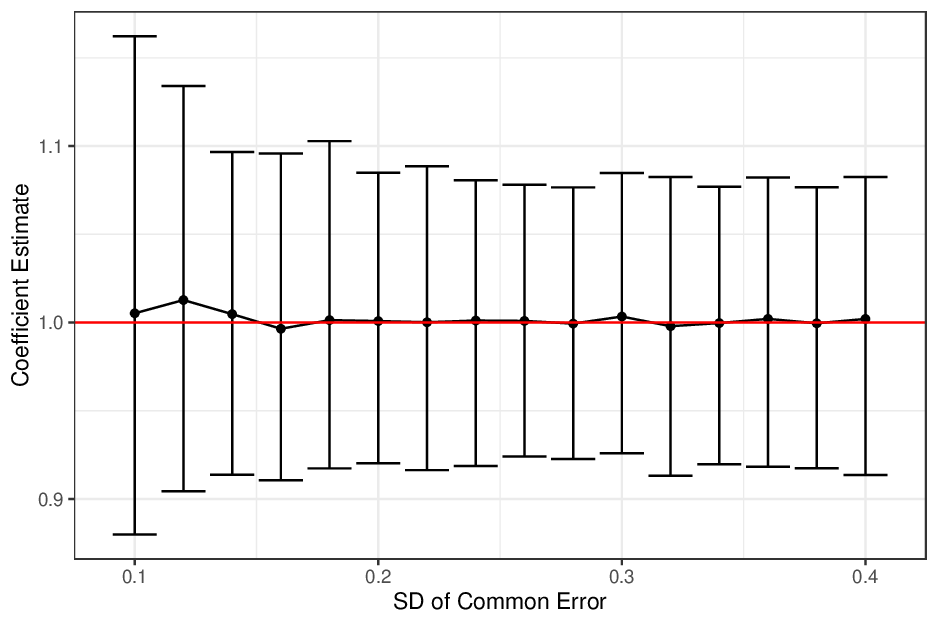}
    \caption{Estimation Results of Extended EnsembleIV (red line indicates true coefficient value; results obtained over 500 repetitions)}
    \label{fig:diagnostic_extension}
\end{figure}

We apply this approach to the aforementioned simulation. We specifically focus on $\sigma \in \{0.1, 0.12, \ldots, 0.4\}$ because they represent situations where our original (unmodified) IV transformation step fails to create valid IVs (as shown in the first plot of Figure \ref{fig:diagnostic_sigma}). Under each value of $\sigma$, we simulate a sample of 14,000 data points and randomly partition it into 3,000 as $D_{train}$, 1,000 as $D_{test}$, and 10,000 as $D_{unlabel}$. As before, we treat $X_1$ as the endogenous variable and $X_2$ as the candidate IV. We use $D_{train}$ and $D_{test}$ to estimate the modified $\lambda$, then carry out IV transformation and estimation on $D_{unlabel}$. In Figure \ref{fig:diagnostic_extension}, we present the mean and empirical 95\% confidence interval of the IV estimate (implementing 500 repetitions).

Compared with the third plot in Figure \ref{fig:diagnostic_sigma}, it is clear that the extended EnsembleIV approach can recover empirically unbiased coefficient estimates in the presence of the peripheral feature challenge, indicating that the modified IV transformation step is able to remove the correlations between prediction error and peripheral feature.

\section{Statistical Properties} \label{ap:stats_theory}
We will now present the asymptotic properties of our EnsembleIV estimator. We begin by establishing the scenario and available data. There are two sets of data: $D_{label} = \{\left(Y_i, X_i, V_i\right)\}_{i=1\ldots n_{l}}$ and $D_{unlabel} = \{\left(Y_i, \cdot,V_i\right)\}_{j=1\ldots n_{u}}$, where $V_i$ represents a vector of features (e.g., text); $X_i = f(V_i)$ is a property of the features (e.g., textual sentiment), which is missing for all elements in $D_{unlabel}$; $Y_i$ is some outcome of interest, and each tuple of data is drawn i.i.d. from some population of interest. Further, $D_{label}$ is partitioned into two subsets: $D_{test}= \{\left(X_i, V_i\right)\}_{i=1\ldots n_{t}}$ and $D_{train}= \{\left(X_i, V_i\right)\}_{i=n_{t+1} \ldots n_{l}}$, where $|D_{test}| = n_{t}$,  $|D_{train}|= n_{tr} = n_{l}-n_{t}$, and $\psi = \frac{n_{t}}{n_{l}}$ represents the (fixed) proportion of the labeled data allocated to $D_{test}$. Please note that, in the second-phase regression, there may be a set of exogenous control covariates, $\boldsymbol{W}$. We omit them for brevity in the following theoretical discussions, because they are exogenous with respect to the ML-generated covariate.

We first impose a set of common regularity conditions on $f$, the function that describes ML predictions. These conditions ensure $f$ is well-behaved for the eventual purpose of estimation, which puts mild limitations on permissible function classes. Specifically, we impose the following regularity conditions:
\begin{itemize}
    \item $|f(V)|\le C_1$ for some constant $C_1$.
    \item $\supp (f) = \mathbb{S} \subset \mathbb{R}$, where $\mathbb{S}$ is finite.
    \item $V \in \mathbb{S} \text{ almost surely}$.
    \item $ \forall\, V \ \exists\, \mu_{f|V} \ \text{such that} \ \mu_{f|V} \ \text{is constant and} \ f(V) \mid V = \mu_{f|V} \ \text{almost surely}.$
    \item $E\left[f(V)\right] = \mu_{f} \ne 0$.
\end{itemize}

Now we turn our attention to the estimation of $f$ using $D_{train}$. Specifically, there is $\hat{f} = \{{\hat{f}}_j\}_{j=1\ldots M}$ a set of $M>1$ unique models, where each element ${\hat{f}}_j(V)$ is capable of providing a distinct prediction of $X$ given $V$, represented by
\begin{align*}
    \widehat{X}^{(j)} &= {\hat{f}}_j(V)\\
        &= f(V) + \left({\hat{f}}_j(V) - f(V)\right)\\
    &= f(V) + e_j(V).
\end{align*}
While EnsembleIV proposes to construct $\hat{f}$ via an ensemble machine learning method, the use of ensemble learners is not a strict requirement for the application of the theory derived in this section. Instead, we consider the following mild regularity conditions on $e_j(V)$, i.e., the prediction error associated with the $j$-th model:
\begin{itemize}
    \item $E\left[e_j(V)|V\right] = 0$.
    \item $E\left[e_j^2(V)|V\right] = \frac{C_2^{(j)}}{n_{tr}^{\gamma}}$ with $\gamma > 0$ and some set of constant $\{C_2^{(j)}\}_{j=1\ldots M}$.
    \item $E\left[e_j(V)e_k(V)|V\right] = \frac{C_3^{(j,k)}}{n_{tr}^{\nu}}$ with $\nu \ge \gamma$ and some set of constant $\{C_3^{(j,k)}\}_{j,k=1\ldots M}$.
\end{itemize}

Next, we consider the main structural equation of interest. For simplicity, and without loss of generality, let us define $\widehat{X} := \widehat{X}^{(1)}$ and $Z := \widehat{X}^{(2)}$ to be two distinct estimations of $f$, with corresponding errors $e_{\widehat{X}} := e_{1}(V)$ and $e_{Z} := e_{2}(V)$. For clarity, we restate a simplified version of the structural equation from \eqref{eq:regression_true}-\eqref{eq:regression_est} in the main text:
\begin{equation}
    \label{eq:regression_redef}
    \begin{split}
        Y &=  X \beta + \varepsilon \\
    &= \widehat{X} \beta + (\varepsilon - e_{\widehat{X}} \beta) \\
    &= \widehat{X} \beta + u.
    \end{split}
\end{equation}
To reiterate, we have removed the presence of exogenous control variables $\boldsymbol{W}$ for mathematical convenience. As discussed in Section \ref{sec:theory_IV} (adapting to the notations and setup in this part), we make the following standard and mild assumptions about the idiosyncratic error term $\varepsilon$:
\begin{itemize}
    \item [Assumption I:] $E[\varepsilon| X] = 0$.
    \item [Assumption II:] Given $\widehat{X}^{(i)} = X + e_i(V)$, $E[\varepsilon e_i(V) | X]=0, ~ \forall i \in \{1, \ldots, M\}$.
\end{itemize}
Assumption I ensures that, in the absence of measurement error, we are working with a correctly specified regression equation free from other sources of endogeneity, while Assumption II ensures that prediction errors associated with each individual learner do not have a direct impact on outcome $Y$ beyond the covariates already included in the regression. Furthermore, to more clearly characterize the various sources of error in the EnsembleIV estimator in finite samples, and to facilitate the demonstration of its asymptotic behavior, we also make the following homoscedasticity assumption:
\begin{itemize}
    \item [Assumption III:] $E\left[\varepsilon^2| X \right] = \sigma^2_{\varepsilon} \le C_4$ for some constant $C_4$.
\end{itemize}
That is, in the absence of measurement error, the variance of $\varepsilon$ conditioned on $X$ is homogeneous and finite. This assumption is notably standard practice in the econometrics literature, resulting in simpler and more intuitive empirical and theoretical modeling. Regression estimation with infinite variance is extremely challenging as most standard estimators rely on various smoothness properties of the underlying data-generating process, which are violated in the presence of infinite variance. Homogeneous variance (i.e., homoscedasticity) is also commonly assumed as it enables more reliable (unbiased and consistent) estimation of the regression coefficient standard errors and, thus, inference on the coefficients themselves.\footnote{Note that this homoscedasticity assumption is not strictly necessary for EnsembleIV estimation and inference because we bootstrap our standard errors.}

Our focal context occurs when the estimation of $\beta$ in \eqref{eq:regression_redef} is carried out on $D_{unlabel}$ with $\widehat{X}$ in place of the unobserved $X$. The implicit objective is to leverage the larger sample size of $D_{unlabel}$ for estimation, given $n_u >> n_l$. However, standard estimation of $\beta$ by Ordinary Least Squares (OLS) using $D_{unlabel}$ in \eqref{eq:regression_redef} is generally biased and inconsistent because machine learning models generally produce imperfect predictions of $X$, and the resulting measurement error yields unreliable inference on $\beta$.

To move forward, despite the imperfect predictions of machine learning models, EnsembleIV introduces the quantity $\lambda$, which is used to build an instrumental variable (IV) for $\widehat{X}$. We restate $\lambda$ with slight notational adjustments from \eqref{eq:lambda} in the main text:
\begin{equation}
  \label{eq:lambda_redef}
  \begin{split}
    \lambda &= \frac{\rho_{Zu}}{\rho_{\widehat{X}u}} \\
    &= \frac{\text{Cov}(Z,u)}{\text{Cov}(\widehat{X},u)} \cdot \frac{\sigma_{\widehat{X}}}{\sigma_Z}\\
    &= \frac{\text{Cov}(Z,e_{\widehat{X}})}{\text{Cov}(\widehat{X},e_{\widehat{X}})} \cdot \frac{\sigma_{\widehat{X}}}{\sigma_Z},
\end{split}
\end{equation}
where the final equality follows from the application of our above assumptions. Moreover, we can also consider a plug-in estimator of $\lambda$: 
\begin{equation}
    \label{eq:lambda_hat_redef}
    \hat{\lambda} = \frac{\widehat{\text{Cov}}(Z,e_{\widehat{X}})}{\widehat{\text{Cov}}(\widehat{X},e_{\widehat{X}})} \cdot \frac{\hat{\sigma}_{\widehat{X}}}{\hat{\sigma}_Z},
\end{equation}
substituting each theoretical $\text{Cov}(\cdot, \cdot)$ and $\sigma$ with an empirical analog $\widehat{\text{Cov}}(\cdot, \cdot)$ and $\hat{\sigma}$ respectively, computed from an available sample of data, where all the necessary variables are observed. Having defined our setting, assumptions, and relevant quantities, we will now begin establishing the theoretical results that enable EnsembleIV, the first of which will focus on the properties of $\lambda$ and $\hat{\lambda}$. Note that all theoretical results rely on Assumptions 1-11 implicitly.

\begin{lemma} \label{lem:lambda2}
   Let $\widetilde{Z} = \sigma_{\widehat{X}} Z - \lambda \sigma_Z \widehat{X}$, then $\text{Cov}(\widetilde{Z}, u)=0$.
\end{lemma}

\proof{Proof.}
This result follows directly from the definition of $\lambda$:
\begin{align*}
\text{Cov}(\widetilde{Z}, u) &= \sigma_{\widehat{X}} \text{Cov}(Z,u) - \frac{\text{Cov}(Z,u)}{\text{Cov}(\widehat{X},u)} \cdot \frac{\sigma_{\widehat{X}}}{\sigma_Z} \sigma_Z \text{Cov}(\widehat{X}, u)\\
&= 0.
\end{align*}
\endproof

\begin{corollary}
\label{cor:hat_lambda}
Given a sample of data elements $D_n = \left(\bm{\widehat{X}}, \bm{Z}, \bm{u}\right) = \left\{\left(\widehat{X}, Z_i, u_i\right)\right\}_{i=1\ldots n}$, $\widetilde{Z}_i = \hat{\sigma}_{\widehat{X_i}} Z_i - \hat{\lambda}_n \hat{\sigma}_Z \widehat{X}_i$, $\bm{\widetilde{Z}} = \{ \widetilde{Z}_i \}_{i=1\ldots n}$, let 
\begin{align*}
    \hat{\lambda}_n &= \frac{\frac{1}{n-1} \sum_{i=1}^{n} (Z_i - \bar{Z})(u_i - \bar{u})}{\frac{1}{n-1} \sum_{i=1}^{n} (\widehat{X}_i - \bar{\widehat{X}})(u_i - \bar{u})} \cdot \frac{\sqrt{\frac{1}{n-1} \sum_{i=1}^{n} (\widehat{X}_i - \bar{\widehat{X}})^2}}{\sqrt{\frac{1}{n-1} \sum_{i=1}^{n} (Z_i - \bar{Z})^2}} \\
    &= \frac{\widehat{\text{Cov}}(\bm{Z},\bm{u})}{\widehat{\text{Cov}}(\bm{\widehat{X}},\bm{u})} \cdot \frac{\hat{\sigma}_{\widehat{X}}}{\hat{\sigma}_Z},
\end{align*}
then 
\begin{equation*}
    \widehat{\text{Cov}}(\bm{\widetilde{Z}}, \bm{u})=0.
\end{equation*}
\end{corollary}

From Lemma \ref{lem:lambda}, we have that the population level transformation of the potentially endogenous instrument $Z$ provides a new instrument, $\widetilde{Z}$, that theoretically meets the exclusion condition. Corollary \ref{cor:hat_lambda} extends this result to a fixed dataset. More specifically, it establishes that transforming the sample vector $\bm{Z}$ by the $\hat{\lambda}_n$ computed on the dataset, provides a new sample vector $\bm{\widetilde{Z}}$ that meets the exclusion condition for the given dataset. While it is possible to compute $\hat{\lambda}_n$ on $D_{test}$ (i.e., $\hat{\lambda}_{n{t}}$), it is generally not possible to compute $\hat{\lambda}_{n{u}}$ on $D_{unlabel}$. And, given $n_{t}>> n_{u}$, our goal is to leverage $D_{unlabel}$ for inference. Therefore, we next analyze the relationship between $\hat{\lambda}_{n_{t}}$ and $\hat{\lambda}_{n_{u}}$.

\begin{lemma}
\label{lem:lambda_convergence}
Let $\hat{\lambda}_{n_{t}}$ be the (computable) plug-in estimator for $\lambda$ based on $D_{test}$, and $\hat{\lambda}_{n_{u}}$ be the (uncomputable) plug-in estimator for $\lambda$ based on $D_{unlabel}$, then $\hat{\lambda}_{n_{t}} \xrightarrow{a.s.} \hat{\lambda}_{n_{u}} ~\text{as}~ n_u, n_t \to \infty$.
\end{lemma}
\proof{Proof.}
We start by defining useful quantities:
\begin{gather*}
    \theta_{n_{tr}} = \left(\text{Cov}(\bm{Z},\bm{e_{\widehat{X}}}), \text{Cov}(\bm{\widehat{X}},\bm{e_{\widehat{X}}}), \sigma_{\widehat{X}}, \sigma_Z\right), \\
    \widehat{\theta}_{n_{tr}, n} = \left(\widehat{\text{Cov}}_n(\bm{Z},\bm{e_{\widehat{X}}}), \widehat{\text{Cov}}_n(\bm{\widehat{X}},\bm{e_{\widehat{X}}}), \hat{\sigma}_{\widehat{X},n}, \hat{\sigma}_{Z,n}\right), \\
    g\left(\left(a,b,c,d\right)\right) = \frac{a}{b} \cdot \frac{c}{d}.
\end{gather*}
Note that $g$ is continuous everywhere except where $b$ or $d$ are zero. Further, we define $\lambda_{n_{tr}}$, making explicit its reliance on the training data sample size $n_{tr}$:
\begin{equation}
  \label{eq:lambda_asym}
  \begin{split}
    \lambda_{n_{tr}} &= g(\theta_{n_{tr}})\\
    &= \frac{\text{Cov}(Z,e_{\widehat{X}})}{\text{Cov}(\widehat{X},e_{\widehat{X}})} \cdot \frac{\sigma_{\widehat{X}}}{\sigma_Z}\\
    &= \frac{\text{Cov}(Z,e_{\widehat{X}})}{\text{Var}(e_{\widehat{X}})} \cdot \frac{\sqrt{\text{Var}(f(V)+e_{\widehat{X}})}}{\sqrt{\text{Var}(f(V)+e_{Z})}}\\
    &= \frac{\Theta({n_{tr}}^{-\nu})}{\Theta({n_{tr}}^{-\gamma})} \cdot \frac{\sqrt{\text{Var}(f(V)+\Theta({n_{tr}}^{-\nu})}}{\sqrt{\text{Var}(f(V)+\Theta({n_{tr}}^{-\nu}))}}\\
    &= \Theta({n_{tr}}^{-\alpha})\quad \text{where } \alpha = \nu-\gamma \ge 0.
    \end{split}
\end{equation}
Therefore, $\lambda_{n_{tr}}$ converges to some constant (specifically zero if $\alpha > 0$).  We also define $\hat{\lambda}_{n_{tr},n}$, making explicit its similar reliance on $n_{tr}$ and additional reliance on the size ($n$) of a separate data sample used for plug-in estimation: 
\begin{equation}
  \label{eq:lambda_hat_converg}
  \begin{split}
    \hat{\lambda}_{n_{tr},n} &= g(\widehat{\theta}_{n_{tr}, n})\\
    &= \frac{\widehat{\text{Cov}}_n(Z,e_{\widehat{X}})}{\widehat{\text{Cov}}_n(\widehat{X},e_{\widehat{X}})} \cdot \frac{\hat{\sigma}_{\widehat{X},n}}{\hat{\sigma}_{Z,n}}\\
    &\overset{\text{a.s.}}{\underset{n \to \infty}{\longrightarrow}}  \frac{\text{Cov}(Z,e_{\widehat{X}})}{\text{Cov}(\widehat{X},e_{\widehat{X}})} \cdot \frac{\sigma_{\widehat{X}}}{\sigma_Z}\\
    &= g(\theta_{n_{tr}})\\
    &=\lambda_{n_{tr}}.
    \end{split}
\end{equation}
Note that \eqref{eq:lambda_hat_converg} follows from $\widehat{\theta}_{n_{tr}, n} \overset{\text{a.s.}}{\underset{n \to \infty}{\longrightarrow}} \theta_{n_{tr}}$ and therefore $g(\widehat{\theta}_{n_{tr}, n}) \overset{\text{a.s.}}{\underset{n \to \infty}{\longrightarrow}}  g(\theta_{n_{tr}})$, by the Strong law of large numbers and the Continuous mapping theorem. Importantly, \eqref{eq:lambda_asym} ensures that \eqref{eq:lambda_hat_converg} holds for any value of $n_{tr}$, including $n_{tr} \rightarrow \infty$ where $\text{Cov}(\widehat{X},e_{\widehat{X}}) \rightarrow 0$.

Finally, we have 
\begin{align*}
|\hat{\lambda}_{n_{tr},n_{t}} - \hat{\lambda}_{n_{tr},n_{u}}| &\leq |\hat{\lambda}_{n_{tr},n_{t}} - \lambda_{n_{tr}}| + |\hat{\lambda}_{n_u} - \lambda_{n_{tr}}|\\
&\overset{\text{a.s.}}{\underset{n_{t},n_{u} \to \infty}{\longrightarrow}} 0,
\end{align*}
by the Triangle inequality and the application of \eqref{eq:lambda_hat_converg} for both $\hat{\lambda}_{n_{tr},n_{t}}$ and $\hat{\lambda}_{n_{tr},n_{u}}$.


Lemma \ref{lem:lambda_convergence}, demonstrates that $\hat{\lambda}_{n_{t}}$ is actually converging almost-surely to $\hat{\lambda}_{n_{u}}$, allowing us to use $\hat{\lambda}_{n_{t}}$ in its place, asymptotically. We can therefore obtain the asymptotic behavior of the EnsembleIV estimator on unlabeled data $D_{unlabel}$, by first analyzing its behavior without the transformation of $Z$ by $\hat{\lambda}_{n_{t}}$
\begin{theorem}
    \label{thm:B_FIV}
    Let $\hat{\beta}_{IV}$ be an instrumental variable estimator of $\beta$ in \eqref{eq:regression_redef} estimated on unlabeled data $D_{unlabel}$, where the instrument $Z$, is a variable with prediction error that asymptotically meets the exclusion condition. Under Assumptions 1-11, as $n_{u},n_{l} \to \infty$

\begin{equation*}
\sqrt{n_u} (\hat{\beta}_{IV} - \beta) \rightsquigarrow \mathcal{N}\left(0, \frac{\sigma^2_{\epsilon}}{\bar{f^2}}\right),    
\end{equation*}

where $\bar{f^2} = \frac{1}{n_u} \sum_{i} f(V_i)^2$.
\end{theorem}
\proof{Proof.}
The IV estimator is defined as:
\begin{align}
    \hat{\beta}_{IV} &= \left( \frac{1}{n_u} \sum_{i=1}^{n_u} Z_i \widehat{X}_i \right)^{-1} \left( \frac{1}{n_u} \sum_{i=1}^{n_u} Z_i Y_i \right) \nonumber \\
    &= \left( \frac{1}{n_u} \sum_{i=1}^{n_u} Z_i \widehat{X}_i \right)^{-1} \left( \frac{1}{n_u} \sum_{i=1}^{n_u} Z_i \left(X_i \beta + \varepsilon_i\right) \right) \label{eq:FIV_define}.
\end{align}
As is common in instrumental variable analysis, we are specifically interested in the behavior of $\hat{\beta}_{IV}$ given a data sample, therefore taking $V$ to be observed and fixed. For notational simplicity, we will leave conditioning on $V$ implicit, except when it proves particularly fruitful to be explicit. We can now proceed by recognizing that the asymptotic bias and variance of $\hat{\beta}_{IV}$ will be dictated by the limit of the right summand in \eqref{eq:FIV_define} and separately analyzing each component. Considering the first component we have
\begin{align*}
\frac{1}{n_u} \sum_{i=1}^{n_u}  Z_i \widehat{X}_i &= \frac{1}{n_u} \sum_{i} \left( f(V_i) + e_2(V_i) \right) \left( f(V_i) + e_1(V_i) \right)\\
&= \frac{1}{n_u} \sum_{i}{ f(V_i)^2 + \left(e_2(V_i) + e_1(V_i)\right) f(V_i) + \left(e_2(V_i) e_1(V_i)\right)},
\end{align*}
with the following first central moment,
\begin{align*}
E\left[\frac{1}{n_u} \sum_{i=1}^{n_u} Z_i \widehat{X}_i  \right]&= E\left[ \frac{1}{n_u} \sum_{i} f(V_i)^2 \right] + E\left[ \frac{1}{n_u} \sum_{i} \left(e_2(V_i) + e_1(V_i)\right) f(V_i) \right] + E\left[\frac{1}{n_u} \sum_{i} \left(e_2(V_i)e_1(V_i)\right)  \right]\\
&= \bar{f^2} + 0 + O\left(\frac{1}{n_{l}^{\nu}} \right)\\
&= \bar{f^2}  + O\left(\frac{1}{n_{l}^{\nu}} \right),
\end{align*}
 and second central moment
\begin{align*}
 \text{Var}\left[ \frac{1}{n_u} \sum_{i=1}^{n_u} Z_i \widehat{X}_i\right] &= \frac{1}{n_u^2} \sum_{i=1}^{n_u} \text{Var}\left[ f(V_i)^2 + \left( e_2(V_i) + e_1(V_i) \right) f(V_i) + e_2(V_i) e_1(V_i) \right] \\
    &= \frac{1}{n_u^2} \sum_{i=1}^{n_u} \Bigg[ \text{Var}\left[ f(V_i)^2 \right] + \text{Var}\left[ \left( e_2(V_i) + e_1(V_i) \right) f(V_i) \right] + \text{Var}\left[ e_2(V_i) e_1(V_i) \right] \\
    &\quad + 2 \cdot \text{Cov}\left( f(V_i)^2, \left( e_2(V_i) + e_1(V_i) \right) f(V_i) \right) + 2 \cdot \text{Cov}\left( f(V_i)^2, e_2(V_i) e_1(V_i) \right) \\
    &\quad + 2 \cdot \text{Cov}\left( \left( e_2(V_i) + e_1(V_i) \right) f(V_i), e_2(V_i) e_1(V_i) \right) \Bigg] \\
    &= \frac{1}{n_u^2} \sum_{i=1}^{n_u} \Bigg[0+ O\left( \frac{1}{n_{l}^{\gamma}} \right) + O\left( \frac{1}{n_{l}^{2\nu}} \right) + 0 + 0 + O\left( \frac{1}{n_{l}^{\nu}} \right)\Bigg]\\
    &= O\left(  \frac{1}{n_{u}} \frac{1}{n_{l}^{\gamma}} \right).
\end{align*}
Now considering the second component we have 
\begin{align*}
\frac{1}{n_u} \sum_{i=1}^{n_u} Z_i Y_i &= \frac{1}{n_u} \sum_{i=1}^{n_u} \left( Z_i X_i \beta + Z_i \varepsilon_i \right)\\
&= \frac{1}{n_u} \sum_{i=1}^{n_u} \left( (f(V_i) + e_2(V_i)) f(V_i) \beta + (f(V_i) + e_2(V_i)) \varepsilon_i \right)\\
&= \frac{1}{n_u} \sum_{i=1}^{n_u} f(V_i)^2 \beta + f(V_i) e_2(V_i) \beta + f(V_i) \varepsilon_i + e_2(V_i) \varepsilon_i
\end{align*}
with the following first central moment
\begin{align*}
E\left[ \frac{1}{n_u} \sum_{i=1}^{n_u} Z_i Y_i \right]  &= \beta E\left[ \frac{1}{n_u} \sum_{i} f(V_i)^2 \right] + \beta E\left[ \frac{1}{n_u} \sum_{i} f(V_i) e_2(V_i)\right] + E\left[\frac{1}{n_u} \sum_{i} 
 f(V_i) \varepsilon_i \right] + \beta E\left[ \frac{1}{n_u} \sum_{i} e_2(V_i) \varepsilon_i\right]\\
&= \beta \bar{f^2}  + 0 + 0 + 0\\
&= \beta \bar{f^2} 
\end{align*}
and second central moment
\begin{align*}
 \text{Var}\left[ \frac{1}{n_u} \sum_{i=1}^{n_u} Z_i Y_i\right]
    &= \text{Var}\left[\frac{1}{n_u} \sum_{i=1}^{n_u} f(V_i)^2 \beta + f(V_i) e_2(V_i) \beta + f(V_i) \varepsilon_i + e_2(V_i) \varepsilon_i  \right]\\
    &= \frac{1}{n_u^2} \sum_{i=1}^{n_u} \text{Var}\left[ f(V_i)^2 \beta \right] + \text{Var}\left[ f(V_i) e_2(V_i) \beta \right] + \text{Var}\left[ f(V_i) \varepsilon_i \right] + \text{Var}\left[ e_2(V_i) \varepsilon_i \right]\\
    &= \frac{1}{n_u^2} \sum_{i=1}^{n_u} \Bigg[0 + O\left( \frac{1}{n_{l}^{\gamma}} \right) + f(V_i)^2 \text{Var}\left[\varepsilon_i | V_i \right] + O\left( \frac{1}{n_{l}^{\gamma}} \right)\Bigg]\\
    &= \frac{\bar{f^2} \sigma^2_{\varepsilon}}{n_u} + O\left( \frac{1}{n_u}\frac{1}{n_{l}^{\gamma}} \right).
\end{align*}

We can therefore use the results of these two components to finalize our analysis of $\hat{\beta}_{IV}$ from \eqref{eq:FIV_define} with 
\begin{align*}
\hat{\beta}_{IV} &= \frac{ \beta \bar{f^2}  + \frac{1}{n_u} \sum_{i} 
 f(V_i) \varepsilon_i+ O_p\left(\frac{1}{\sqrt{n_{u}}} \frac{1}{n_{l}^{\frac{\gamma}{2}}}\right)}{\bar{f^2}  + O\left(\frac{1}{n_{l}^{\nu}} \right) +O_p\left(\frac{1}{\sqrt{n_{u}}} \frac{1}{n_{l}^{\frac{\gamma}{2}}} \right)}.
\end{align*}

Moreover, we can define $Z_{IV}$ as follows
\begin{align*}
    Z_{IV} &=\sqrt{n_u} (\hat{\beta}_{IV} - \beta)\\
            &=  \sqrt{n_u}\left(\frac{ \beta \bar{f^2}  + \frac{1}{n_u} \sum_{i} f(V_i) \varepsilon_i+ O_p\left(\frac{1}{\sqrt{n_{u}}} \frac{1}{n_{l}^{\frac{\gamma}{2}}}\right)}{\bar{f^2}  + O\left(\frac{1}{n_{l}^{\nu}} \right) +O_p\left(\frac{1}{\sqrt{n_{u}}} \frac{1}{n_{l}^{\frac{\gamma}{2}}} \right)} - \beta \right) \\
            &=  \sqrt{n_u} \beta \left(\frac{\bar{f^2}}{\bar{f^2}  + O\left(\frac{1}{n_{l}^{\nu}} \right) +O_p\left(\frac{1}{\sqrt{n_{u}}} \frac{1}{n_{l}^{\frac{\gamma}{2}}} \right)} -1  \right) + \frac{\frac{1}{\sqrt{n_u}} \sum_{i} f(V_i) \varepsilon_i+ O_p\left(\frac{1}{n_{l}^{\frac{\gamma}{2}}}\right)}{\bar{f^2}  + O\left(\frac{1}{n_{l}^{\nu}} \right) +O_p\left(\frac{1}{\sqrt{n_{u}}} \frac{1}{n_{l}^{\frac{\gamma}{2}}} \right)}\\
            &=  \sqrt{n_u} \beta \left(\frac{\bar{f^2}}{\bar{f^2}  + o_p\left(1\right)} -1  \right) + \frac{\frac{1}{\sqrt{n_u}} \sum_{i} f(V_i) \varepsilon_i+ o_p\left(1\right)}{\bar{f^2}  + o_p\left(1\right)}\\
            &\underset{n_{l},n_{u} \to \infty}{\rightsquigarrow}  \mathcal{N}\left(E\left[\frac{\frac{1}{\sqrt{n_u}} \sum_{i} f(V_i) \varepsilon_i}{\bar{f^2}}\right], \text{Var}\left[\frac{\frac{1}{\sqrt{n_u}} \sum_{i} f(V_i) \varepsilon_i}{\bar{f^2}}\right]\right)\\
            &= \mathcal{N}\left(0, \frac{\sigma^2_{\varepsilon}}{\bar{f^2}}\right),
\end{align*}
where the convergence in distribution is implied by convergence in probability and Slutsky's theorem.

\begin{corollary}
\label{cor:B_EIV}
Let $\widetilde{Z}_i = \hat{\sigma}_{\widehat{X}} Z_i - \hat{\lambda}_{n_{t}} \hat{\sigma}_Z \widehat{X}_i$ be the transformation of each $Z_i$ in $D_{unlabel}$ and let $\hat{\beta}_{EIV}$ be an estimator with instrument $\widetilde{Z}$ as defined in Theorem \ref{thm:B_FIV}, then by Corollary \ref{cor:hat_lambda} and Lemma \ref{lem:lambda_convergence}, as $n_{u},n_{l} \to \infty$
\begin{equation*}
\sqrt{n_u} (\hat{\beta}_{EIV} - \beta) \rightsquigarrow \mathcal{N}\left(0, \frac{\sigma^2_{\epsilon}}{\bar{f^2}}\right).
\end{equation*}
The specific rate of convergence experienced by $\hat{\beta}_{EIV}$ is dictated by $\gamma$ and $\nu$, which are governed by the convergence rate of $\hat{\lambda}_{n_{t}}$ to $\hat{\lambda}_{n_{u}}$.
\end{corollary}

\underline{\textbf{Remark:}} The challenge of conducting inference after the use of flexible machine learning algorithms for model selection (i.e, post-selection inference) has been considered extensively \citep{berk2013valid,belloni2014post-inference,chernozhukov2015post-inference,belloni2016post-inference-glm,belloni2017high-inference}. One critical challenge this literature attempts to address occurs when the outcome model in which inference is being carried out is selected in a data-driven manner \citep{berk2013valid}. A second challenge occurs when the variable of inferential interest (e.g., treatment) in the outcome model is not observed exogenously, and therefore the treatment and the outcome are jointly determined by a subset of variables. Data-dependent selection for these control variables can create a type of ``regularization bias" by incorrectly estimating the functional form with which the control variables enter the treatment or the outcome models, including failing to include relevant controls \citep{chernozhukov2015post-inference}. Our setting described in Equation \eqref{eq:regression_true} assumes an outcome model where $E[\varepsilon| X, \boldsymbol{W}] = 0$ and the variables used to build $X$ (e.g., textual features that determine sentiment) have no direct impact on the outcome conditional on $X$. Therefore, data-driven selection of the treatment model (i.e., the model of $X$) will not lead to estimation bias of $\beta$ in the outcome model. Essentially, in Equation \eqref{eq:regression_true} $X$ is exogenous by assumption, hence the only potential source of bias in Equation \eqref{eq:regression_est} arises from using $\widehat{X}$ in place of $X$ due to prediction errors. There may exist contexts where these assumptions do not hold, and the extensive and focused statistical innovations required to relax these assumptions would be of future interest.

\section{Using Subset of Individual Learners as Endogenous Covariates and IVs} \label{ap:subset}
On the Bike Sharing dataset, we first built a random forest of 100 trees and then constructed predictions based on subsets of 50 trees. Among all $\binom{100}{50}$ different ways to select 50 trees out of 100, we randomly sampled 100, to maintain the same number of covariates as in EnsembleIV with individual trees, enabling an apples-to-apples comparison. The estimation results are reported in Table \ref{table:subset}, below. We can see that, compared to using individual trees as covariates, using subsets of 50 trees results in slightly more biased point estimates, inflated standard errors, and overall worse estimation MSE scores.

\begin{table}[!tbh]
    \centering 
    \caption{EnsembleIV Estimates with Predictions from Subsets of Trees} \label{table:subset} 
    \begin{tabular}{c c | c c c || c c c}
    \hline
    &   & \multicolumn{3}{c||}{Individual Trees} & \multicolumn{3}{c}{Subsets of 50 Trees} \\
    \hline
    & True & Top-3 & PCA & LASSO & Top-3 & PCA & LASSO \\ 
    \hline
    $\beta_0$       & 1.0 & 1.026 & 1.015 & 1.058 & 0.956 & 0.941 & 0.903 \\ 
                    &     & (0.063) & (0.063) & (0.062) & (0.069) & (0.073) & (0.073) \\ 
    $\beta_{MLV}$   & 0.5 & 0.494 & 0.496 & 0.487 & 0.510 & 0.513 & 0.521 \\ 
                    &     & (0.013) & (0.013) & (0.013) & (0.015) & (0.015) & (0.016) \\ 
    $\beta_{W_1}$   & 2.0 & 2.000 & 2.000 & 2.000 & 2.000 & 2.000 & 2.000 \\ 
                    &     & (0.003) & (0.003) & (0.003) & (0.003) & (0.003) & (0.003) \\ 
    $\beta_{W_2}$   & 1.0 & 1.000 & 1.000 & 1.000 & 1.000 & 1.000 & 1.000 \\ 
                    &     & (0.002) & (0.002) & (0.002) & (0.001) & (0.001) & (0.001) \\ 
    \hline
    Estimation MSE             &     & 0.005 & 0.004 & 0.008 & 0.007 & 0.009 & 0.015 \\
    \hline
    \end{tabular}
\end{table}

\end{document}